\documentclass[reprint,superscriptaddress,amsmath,amssymb,aps,pre]{revtex4-1}

\usepackage{graphicx}
\usepackage{dcolumn}
\usepackage{bm}
\usepackage{color}
\usepackage{bbold}
\usepackage{soul}

\usepackage[margin=8mm,textfont=small,
position=top,
labelformat=empty,
caption=false,
justification=raggedright, singlelinecheck=false]{subfig}

\newcommand{\ud}{\text{d}}
\providecommand*{\oneD}{\textsc{1d}}
\providecommand*{\twoD}{\textsc{2d}}
\providecommand*{\fourD}{\textsc{4d}}

\providecommand*{\threeD}{\textsc{3d}}
\providecommand*{\threeDD}{\textsc{3D}}
\providecommand*{\fiveD}{\textsc{5d}}
\providecommand*{\sixD}{\textsc{6d}}

\providecommand*{\ui}{\text{i}}

\providecommand*{\valfixpt}[1]{\ensuremath{{\nu_{#1}^\fp}}}
\providecommand*{\nua}{\ensuremath{\valfixpt{x}}}
\providecommand*{\nub}{\ensuremath{\valfixpt{y}}}

\providecommand*{\Mone}{\ensuremath{{\cal M}_x^{\fp}}}
\providecommand*{\Mtwo}{\ensuremath{{\cal M}_y^{\fp}}}

\providecommand*{\fp}{\ensuremath{\mathrm{fp}}}
\providecommand*{\pointPSname}{\ensuremath{u}}

\providecommand*{\pointPS}{\ensuremath{\mathbf{\pointPSname}}}
\providecommand*{\fixedpoint}{\ensuremath{{\pointPS_{\fp}}}}
\providecommand*{\fixedpointtwod}{\ensuremath{{\pointPS_{\fp}^{\twoD}}}}

\providecommand*{\ps}{phase space}
\providecommand*{\confs}{configuration space}
\providecommand*{\pss}{\threeD{} phase-space slice}

\providecommand*{\poincare}{Poincar\'e}
\providecommand*{\poinsec}{\poincare{} section}

\providecommand*{\Prt}{\poincare{} recurrence theorem}
\providecommand*{\Prs}{\poincare{} recurrence statistics}

\providecommand*{\pol}{power-law}

\providecommand*{\parab}{\( \partial \Omega_1 \)}
\providecommand*{\ground}{\( \partial \Omega_2 \)}

\providecommand*{\initregion}{\ensuremath{\Lambda}}

\providecommand*{\onetori}{\oneD{} tori}
\providecommand*{\onetorus}{\oneD{} torus}
\providecommand*{\twotori}{\twoD{} tori}
\providecommand*{\twotorus}{\twoD{} torus}

\providecommand*{\confx}{x}
\providecommand*{\confy}{y}
\providecommand*{\confz}{z}

\providecommand*{\momx}{p_x}
\providecommand*{\momy}{p_y}
\providecommand*{\momz}{p_z}

\newcommand{\bfn}{\hat{\mathbf{n}}}
\newcommand{\qcol}{\mathbf{q}}  

\newcommand{\nuL}{\ensuremath{\nu_{\text{L}}}}
\newcommand{\nuT}{\ensuremath{\nu_{\text{N}}}}

\newcommand{\nuone}{\ensuremath{\nu_{x}}}
\newcommand{\nutwo}{\ensuremath{\nu_{y}}}
\newcommand{\limacon}{lima\c{c}on}

\newcommand{\R}{\mathbb{R}}
\newcommand{\N}{\mathbb{N}}

\usepackage{color}

\usepackage[nice]{nicefrac}

\newcommand{\FBOX}[1]{\fbox{\parbox[center][0.45em][c]{0.45em}{#1}}}

\usepackage{prettyref}
\newrefformat{sec}{Sec.~\ref{#1}}
\newrefformat{app}{Appendix~\ref{#1}}
\newrefformat{fig}{Fig.~\ref{#1}}
\newrefformat{tab}{Tab.~\ref{#1}}
\newrefformat{eq}{Eq.~(\ref{#1})}

\usepackage{catchfile}
\newcommand{\getenv}[2][]{%
\CatchFileEdef{\temp}{"|kpsewhich --var-value #2"}{\endlinechar=-1}%
\if\relax\detokenize{#1}\relax\temp\else\let#1\temp\fi}

\getenv[\articlestoragepath]{ARTICLE_STORAGE}
\IfFileExists{pdf_article_link.sty}{\usepackage{pdf_article_link}}

\newcommand{\MOVIEREF}{%
   For a rotating view see
  \href{http://www.comp-phys.tu-dresden.de/supp/}
          {http://www.comp-phys.tu-dresden.de/supp/}.}
\newcommand{\MOVIELINK}{%
    \href{http://www.comp-phys.tu-dresden.de/supp/}
    {http://www.comp-phys.tu-dresden.de/supp/}}

\makeatletter
\let\Hy@backout\@gobble
\makeatother

\newcommand{\HIDDEN}[1]{}

\begin{document}

\title{3D billiards: visualization of regular structures and trapping
       of chaotic trajectories}

\author{Markus Firmbach}
\affiliation{Technische Universit\"at Dresden, Institut f\"ur Theoretische
             Physik and Center for Dynamics, 01062 Dresden, Germany}
\affiliation{Max-Planck-Institut f\"ur Physik komplexer Systeme, N\"othnitzer
    Stra\ss{}e 38, 01187 Dresden, Germany}

\author{Steffen Lange}
\affiliation{Technische Universit\"at Dresden, Institut f\"ur Theoretische
             Physik and Center for Dynamics, 01062 Dresden, Germany}

\author{Roland Ketzmerick}
\affiliation{Technische Universit\"at Dresden, Institut f\"ur Theoretische
             Physik and Center for Dynamics, 01062 Dresden, Germany}
\affiliation{Max-Planck-Institut f\"ur Physik komplexer Systeme, N\"othnitzer
Stra\ss{}e 38, 01187 Dresden, Germany}

\author{Arnd B\"acker}
\affiliation{Technische Universit\"at Dresden, Institut f\"ur Theoretische
             Physik and Center for Dynamics, 01062 Dresden, Germany}
\affiliation{Max-Planck-Institut f\"ur Physik komplexer Systeme, N\"othnitzer
Stra\ss{}e 38, 01187 Dresden, Germany}

\date{\today}

\begin{abstract}
The dynamics in three-dimensional $(\threeD)$ billiards leads,
using a Poincar\'e section, to a four--dimensional map
which is challenging to visualize.
By means of the recently introduced \pss s
an intuitive representation of the organization of the
mixed phase space with regular and chaotic dynamics is obtained. Of
particular interest for applications are constraints to classical transport
between different regions of phase space
which manifest in the statistics of Poincar\'e recurrence times.
For a \threeD{} paraboloid billiard
we observe a slow power-law decay caused by long-trapped trajectories
which we analyze in phase space and in frequency space.
Consistent with previous results for \fourD{} maps we find that:
(i) Trapping takes place close to regular structures outside the Arnold web.
(ii) Trapping is not due to a generalized island-around-island hierarchy.
(iii) The dynamics of sticky orbits is
governed by resonance channels which extend far into the chaotic sea.
We find clear signatures of partial transport barriers.
Moreover, we visualize the geometry of stochastic layers
in resonance channels explored by sticky orbits.
\end{abstract}

\pacs{PACS here}

\maketitle

\section{Introduction}\label{sec:intro}

Billiard systems are Hamiltonian systems playing
an important role in many areas of physics.
They are given by the free motion of a point particle moving
along straight lines inside some Euclidean domain with
specular reflections at the boundary.
The dynamics is studied in much
detail \cite{KozTre1991,Tab1995,CheMar2006}
and ranges from integrable motion, e.g.\
for billiards in a circle, ellipse or rectangle,
to fully chaotic dynamics, e.g.\ for the
Sinai--billiard \cite{Sin1970},
the Bunimovich stadium billiard \cite{Bun1979},
or the cardioid billiard \cite{Rob1983,Woj1986,Mar1988,BaeDul1997}.

Of particular interest is the generic situation
with a mixed phase space in which regular motion and
chaotic motion coexist \cite{MarMey1974}.
This occurs for example when the billiard is convex
and the boundary is sufficiently smooth,
e.g.\ a slight deformation of the circle
such as the family of \limacon{} billiards \cite{Rob1983, DulBae2001}.
For the class of mushroom billiards a sharply divided mixed phase
space is rigorously proven \cite{Bun2001}.
Billiards also are important model systems in quantum chaos
\cite{Sto2007b,Haa2010}
and have applications in optical microcavities
for which the classical dynamics allows for understanding
and tuning directed laser emission
\cite{GmaCapNarNoeStoFaiSivCho1998,CaoWie2015}.

\begin{figure}[!b]
    \includegraphics{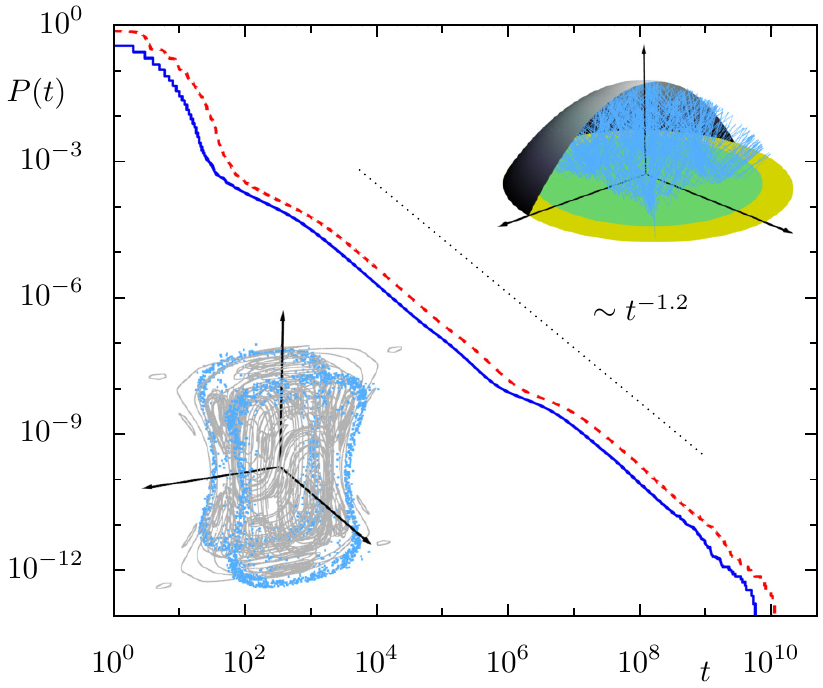}
    \caption{\label{fig:recurrence}
        \Prs{} $P(t)$ in the \threeD{} billiard,
        defined in \prettyref{eq:def_billiard},
        for real
        flight time (red dashed line) and
        number of mappings (blue line).
        The dotted line indicates a \pol{} decay \( \sim t^{-\gamma} \)
        with $\gamma = 1.2$.
        Upper inset:
        \threeD{} paraboloid billiard shown with part of boundary
        removed for visual reasons.
         Inside a sticky trajectory (blue line)
        is shown which starts and returns to
        region \initregion{} (yellow ring).
        Lower inset: trapped orbit (blue dots) and
        regular \ps{} structures (gray) in a \pss{} representation.
}
\end{figure}

Three-dimensional $(\threeD)$ billiards (see the upper right inset in
\prettyref{fig:recurrence}
for an illustration) have in particular been
investigated for establishing fully chaotic dynamics
\cite{Bun1988,Woj1990,BunCasGua1996,BunReh1997,BunReh1998,BunReh1998b,Bun2000,BunMag2006,Woj2007,RapRomTur2007,Sza2017},
and studying both classical and quantum properties
of integrable, mixed and fully chaotic systems, see e.g.\
\cite{ZasStr1992, PriSmi1995,PriSmi2000,AltGraHofRanRehRicSchWir1996,Pro1997a,Pro1997b,Sie1998,Kni1998,WaaWieDul1999,Pap2000,Pap2000b,PapPro2000,DemDieGraHeiPapRicRic2002,DieMoePapReiRic2008,CasRam2011,GilSan2011}.
Recent applications are in the context of
three-dimensional optical micro-cavities
\cite{MekNoeCheStoCha1995, RakYanMcCDonPerMooGapRog2003,
ChiDumFerFerJesNunPelSorRig2010, KreSinHen2017}.
Three-dimensional billiards are also of conceptual interest
because for systems with more than two degrees-of-freedom
new types of transport are possible, including the
famous Arnold diffusion \cite{Arn1964,Chi1979,Loc1999,Dum2014}.

To understand the dynamics of billiards with a mixed
phase space, for \twoD{} billiards the dynamics
in the four--dimensional phase space is conveniently reduced
to a \twoD{} area--preserving map
using energy conservation and a Poincar\'e section.
This can be easily visualized
and used for interactive computer explorations, see e.g.~\cite{Bae2007c}.
In contrast, for \threeD{} billiards the phase space
is six--dimensional and, by energy conservation and a Poincar\'e section,
a \fourD{} symplectic map is obtained, which is difficult to visualize.
One method is to use the recently introduced \pss{} representation
to visualize the regular structures of
\fourD{} symplectic maps \cite{RicLanBaeKet2014},
e.g.\ of two coupled standard maps \cite{Fro1972}.
By this approach it is possible to obtain a good overview
of regular phase space structures and to demonstrate the
generalized island-around-island hierarchy~\cite{LanRicOnkBaeKet2014}
and the organization in terms of families of elliptic
\oneD{} tori~\cite{OnkLanKetBae2016}.

Another motivation comes from the
important question on possible (partial) barriers limiting the transport
between different regions in phase space in higher-dimensional systems.
A sensitive measure for this is the statistics of Poincar\'e recurrence
times $P(t)$. Fully chaotic systems typically show
a fast exponential decay,
see e.g.~\cite{BauBer1990, ZasTip1991,HirSauVai1999, AltSilCal2004},
while for systems with a mixed phase space the decay of $P(t)$ is
much slower, usually following a power-law
\cite{ChaLeb1980,ChiShe1983, Kar1983, ChiShe1984, KayMeiPer1984a,
HanCarMei1985, Mei1986,
MeiOtt1985, MeiOtt1986, ZasEdeNiy1997, BenKasWhiZas1997, ChiShe1999,
ZasEde2000, WeiHufKet2003, CriKet2008,
CedAga2013, AluFisMei2014, AluFisMei2017}.
For recent results on the recurrence time statistics in
integrable systems see \cite{DetMarStr2016}.
Closely related to studying the Poincar\'e recurrence
statistics is the survival probability in open billiards,
see e.g.~\cite{AltTel2008,AltTel2009,AltPorTel2013,DetRah2014}.

For two-dimensional systems the mechanism of the power-law decay of
the Poincar\'e recurrence statistics $P(t)$ is well understood:
here \oneD{} regular tori are absolute barriers to the motion
and thus separate different regions in phase space.
Broken regular tori, so-called cantori,
form partial barriers allowing for a limited transport
\cite{Aub1978, Per1980, BenKad1984, KayMeiPer1984a, KayMeiPer1984b, Wig1990, RomWig1990, RomWig1991, Mei1992, Mei2015}.
Near a regular island formed by invariant
Kolmogorov-Arnold-Moser (KAM) curves,
there is a whole hierarchy associated with
the boundary circle \cite{GreMacSta1986}
and islands-around-islands \cite{Mei1986}.
These hierarchies of partial barriers
are the origin of sticky chaotic trajectories in the surrounding
of a regular island and lead to a power-law behavior of
the Poincar\'e recurrence statistics $P(t)$, see
Refs.~\cite{HanCarMei1985, MeiOtt1985, MeiOtt1986, CriKet2008, AluFisMei2017},
and the reviews~\cite{Mei1992,Mei2015}.

For higher-dimensional systems a power-law decay
of the Poincar\'e recurrence statistics is also commonly observed, see e.g.\
\cite{KanKon1989, KonKan1990, DinBouOtt1990, ChiVec1993, ChiVec1997,
AltKan2007, ShoLiKomTod2007b, She2010, LanBaeKet2016}
and \prettyref{fig:recurrence} for an illustration.
However, an understanding as in the case of two-dimensional
systems is still lacking. The main
reason is that, for example for a \fourD{} map,
the regular tori are two-dimensional and therefore cannot separate different
regions in the \fourD{} phase space. Thus broken regular \twoD{} tori alone
cannot form a partial barrier limiting transport.

In this paper we visualize the dynamics of \threeD{} billiards using the \pss{}
representation and based on this
investigate stickiness of chaotic orbits.
Using the \pss{} reveals how the regular region is organized
around families of elliptic \oneD{} tori and how
uncoupled and coupled resonances govern the regular structures in phase space.
These can be related to trajectories in configuration space
and the representation of the regular region in frequency space.
The Poincar\'e recurrence statistics shows an overall power-law decay.
To investigate this decay, one representative
long-trapped orbit is analyzed in detail
in the \pss{} and in frequency space.
We confirm the findings of Ref.~\cite{LanBaeKet2016} for a \fourD{} map
also in the case of a \threeD{} billiard:
(i) Trapping takes place close to regular structures outside the Arnold web.
(ii) Trapping is not due to a generalized island-around-island hierarchy.
(iii) The dynamics of sticky orbits is
governed by resonance channels which extend far into the chaotic sea.
Clear signatures of partial barriers are found in frequency space
and phase space.
Moreover, we visualize the geometry of stochastic layers
in resonance channels explored by sticky orbits.

This paper is organized as follows:
The first aim is to obtain a visualization of
the mixed phase of a generic \threeD{} billiard.
For this we briefly introduce in Sec.~\ref{sec:billiards} billiard systems,
the \poincare{} section, and as specific example
the \threeD{} paraboloid billiard.
In Sec.~\ref{sec:phasespaceslice} we review and illustrate \pss{}s
and compare with trajectories in configuration space.
The representation in frequency space is discussed in
Sec.~\ref{sec:freqspace}.
The generalized island-around-island hierarchy is discussed
in Sec.~\ref{sec:hierarchyps}
and properties of resonance channels in Sec.~\ref{sec:channels-AD}.
Understanding the transport in a higher-dimensional
system is the second aim of this paper.
For this the Poincar\'e recurrence statistics is introduced
and numerical results
for the \threeD{} paraboloid billiard are presented
in Sec.~\ref{sec:prs}.
The origin of the algebraic decay of the Poincar\'e recurrence statistics
are long-trapped orbits, which we analyze in detail
in Sec.~\ref{sec:long_trapped}
using different representations in phase space and in frequency space.

\section{Visualizing the dynamics of \threeDD{} billiards}

\subsection{Billiard dynamics and \poincare{} section} \label{sec:billiards}

A \threeD{} billiard system  is given as autonomous
Hamiltonian system
\begin{align}
\label{eq:Hamiltonian}
        H\left(\mathbf{p}, \mathbf{q}\right) =   \begin{cases}
    \mathbf{p}^2, & \mathbf{q} \in \Omega \\
    \infty, & \mathbf{q	} \in \partial\Omega \:,\\    \end{cases} \quad
\end{align}
which describes the dynamics of a
freely moving point particle within a closed domain \( \Omega \subset
\mathbb{R}^{3} \) with specular reflections at the boundary
\( \partial\Omega \).
The boundary \( {\partial\Omega = \cup_{i=1}^{n} \partial\Omega_i }\)
is assumed to consist of a finite number of piece-wise smooth
elements $\partial\Omega_i$.
Every point \( \mathbf{q} \in \partial\Omega \) has a  unique
inward pointing
unit normal vector \( \bfn \left(\mathbf{q} \right) \), except for
intersections of boundary elements.

After the free propagation inside the domain the particle collides at a
specific point \( \qcol \in \partial\Omega \) with
the boundary. With respect to the normal vector \( \bfn \left(
  \qcol \right) \) the normal projection of the
momentum vector changes its sign, while the tangent component remains the same.
Therefore the new momentum \( \mathbf{p'} \) after a reflection is given by
\begin{align}
\label{eq:reflection}
    \mathbf{p'}  = \mathbf{p} - 2 \, \left(\mathbf{p} \cdot
       \bfn \left( \qcol \right) \right) \;
    \bfn \left( \qcol \right) ,
\end{align}
where \( \mathbf{p} \) is the momentum before the reflection.

The dynamics of a \threeD{} billiard takes place in a \sixD{} phase
space with coordinates
\( \left( \momx, \momy, \momz, \confx, \confy, \confz \right) \).
As the Hamiltonian \eqref{eq:Hamiltonian}
is time--independent, energy is conserved, i.e.\
$H\left(\mathbf{p}, \mathbf{q} \right)$ is constant
so that the dynamics takes place on a \fiveD{} sub-manifold
of constant energy. As the character of the dynamics
does not depend on the value of $\|\mathbf{p}\|$,
we fix the energy shell by requiring $\|\mathbf{p}\| = 1$.
A further reduction is obtained by introducing
a Poincar\'e section.
This leads to a discrete-time
billiard map on a \fourD{} phase space.
A good parametrization of the section depends on the considered billiard.
Note that for \twoD{} billiards the phase space is four-dimensional and
the whole boundary $\partial \Omega$ usually provides a good section.
Here the section is conveniently parametrized in Birkhoff coordinates
\cite{Bir1966} by the arc-length along the boundary
and the projection of the (unit) momentum vector onto the unit tangent vector
in the point of reflection. In these coordinates one obtains
a \twoD{} area-preserving map \cite{Bir1966}.

As an explicit example we consider the
\threeD{} paraboloid billiard whose domain is defined
by a downwards opened paraboloid \parab{} cut by the plane $z=0$,
leading to an ellipsoid surface as boundary \ground,
\begin{equation}\label{eq:def_billiard}
\begin{aligned}
&\partial\Omega_1 = \left\lbrace \confz = 1 -\frac{1}{2}\left(\left(
\frac{\confx}{a}\right)^2  +
\left(\frac{\confy}{b}\right)^2 \right), \: z \geq 0 \right\rbrace \\
&\partial\Omega_2 = \left\lbrace
\confz=0,  \:
\frac{1}{2}\left(\left(\frac{\confx}{a}\right)^2
+\left(\frac{\confy}{b}\right)^2 \right) \leq 1  \right\rbrace,
\end{aligned}
\end{equation}
with parameters \( a=1.04 \) and \( b=1.12 \).
These parameters are chosen such that
the billiard has no rotational symmetry, $a \neq b$, and
that the central periodic orbit (going along the line $x=0$, $y=0$) is stable
as $a,b > 1$.
The shape of
the system is illustrated in the upper inset in
\prettyref{fig:recurrence} and in \prettyref{fig:slice-vs-trajectories} where
only one half of the paraboloid \parab{} is drawn
and \ground{} is shown in green and yellow.
Note that for $a=b$ the \(z \) component of the angular momentum is conserved.
Numerically this billiard allows for a particularly convenient implementation
as reflection points can be computed by solving a
quadratic equation \cite{Hig2002}.
\begin{figure}[!b]
    \subfloat[][]{
        \includegraphics{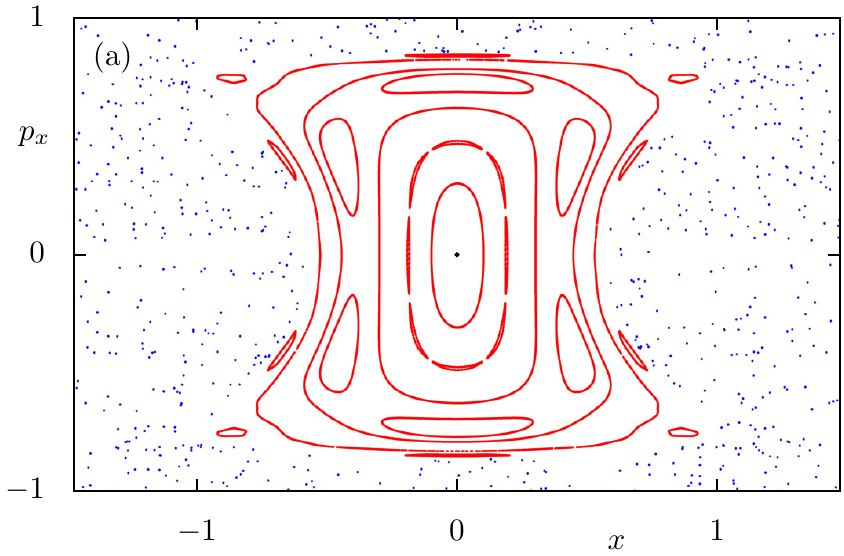}
        \label{fig:ps_2d_04}
    }\\[-2ex]
    \subfloat[][]{
        \includegraphics{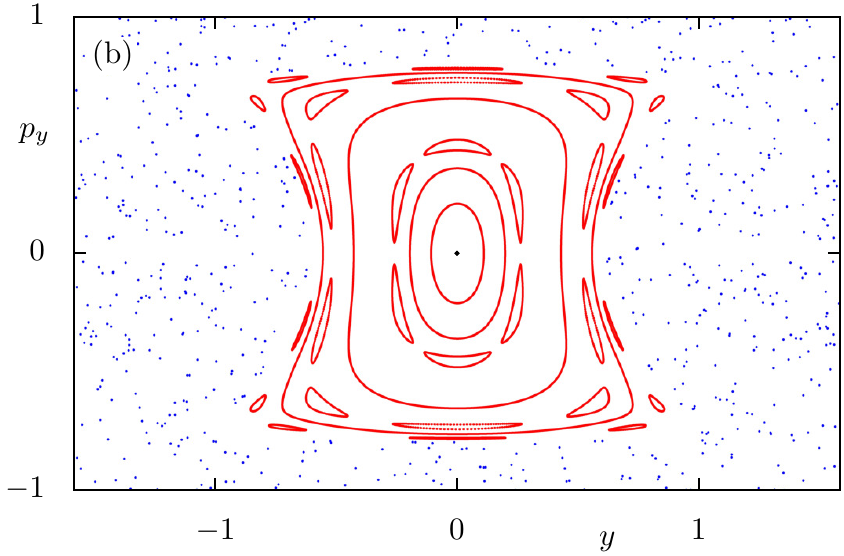}
        \label{fig:ps_2d_12}
    }
    \caption{\label{fig:phasespace2d}%
      Phase space for the two \twoD{} billiards embedded
      in the \threeD{} paraboloid billiard.
      Regular tori (red rings) are
        arranged around an elliptic fixed point (black dot) in
        the center \( {\fixedpointtwod =
        \left(0, 0\right)} \). The regular island
        is embedded in the chaotic region, for which
        one chaotic orbit is shown (blue dots).
        (a) For \( a = 1.04 \) two resonances \( {8\!:\!3} \) and
        \( {3\!:\!1} \) within the regular region and one \( {10\!:\!3} \)
        resonance at the edge of the regular island are shown.
        (b) For \( b = 1.12 \) two resonances \( {3\!:\!1} \) and
        \( {10\!:\!3} \) within the regular region  and one
        \( {14\!:\!4} \) resonance at the edge of the island are shown.
    }
\end{figure}

As \poinsec{} we choose the plane $z=0$,
so that an initial condition is uniquely specified
by its location $(x, y)$ within the ellipse \ground{}
and the momentum components \(\left(\momx, \momy \right) \)
since the third component follows from momentum conservation
$\|\mathbf{p}\| = 1$.
This reduces the \threeDD{} billiard flow
with \sixD{} \ps{} to a symplectic Poincar\'e map \(
\left(\momx, \momy, \confx, \confy \right)\mapsto \left(\momx',
  \momy', \confx', \confy' \right) \)
on a \fourD{} phase space,
\begin{equation}
\begin{aligned}
M =  \{
 & (\momx, \momy, \confx, \confy) \;| \;
      (\confx, \confy) \in \partial \Omega_2,\\
 & (\momx, \momy) \in \R^2 \text{ with } \momx^2 +  \momy^2 \le 1\}
\end{aligned}
\end{equation}
with invariant measure
$\ud \mu =
\frac{1}{|\partial\Omega_2| \pi} \, \ud\momx \ud\momy \ud\confx \ud\confy$.
Note that the trajectory can be
reflected several times at the curved boundary
\parab{} before returning to \ground{}.
There are two different time measures, namely
the number $t$ of applications of the \poincare{} map and the real
flight time $\tau$,
which is the sum of the geometric lengths between consecutive
reflections at the billiard boundary $\partial \Omega$.

Let us first discuss two special cases for the dynamics
of the \threeD{} paraboloid billiard. Corresponding to the motion in the
\mbox{$x$-$z$} and the $y$-$z$~plane
there are two embedded \twoD{} billiards with boundary
given by a straight line and parabola with parameters
$a$ and $b$, respectively.
The central periodic orbit has perpendicular
reflections at the boundaries and geometric length 2.
As $a, b>1$, the radius of curvature is larger
than $ 1 $ and thus larger than half of the length of the periodic orbit
so that in both cases the central periodic orbit is stable.
The phase space of the corresponding billiard maps
in $\left(\confx, \momx \right)$ and  $\left( \confy, \momy \right)$,
respectively, is shown in \prettyref{fig:phasespace2d}.

For the billiard map the stable periodic orbit corresponds
to an elliptic fixed point at the center
${\fixedpointtwod = (q_i, p_i) = (0, 0)}$,
indicated as black dot in \prettyref{fig:phasespace2d}(a,b).
As follows from KAM theory
\cite{Poe2001,Lla2001,Dum2014}
these elliptic points are surrounded by invariant regular tori,
shown as red rings.
Between these KAM tori of sufficiently irrational rotation
frequency one has nonlinear $r:s$ resonance chains,
as implied by the Poincar\'e-Birkhoff theorem \cite{Poi1912,Bir1913},
leading to small embedded sub-islands.
Note that we choose the numbers $r$ and $s$
such that $r$ is the number of sub-islands of a resonance.
The phase space of the billiard map in $\left(\confx, \momx \right)$,
see \prettyref{fig:ps_2d_04}, shows
a prominent \( {6\!:\!2} \) and a smaller
\( {8\!:\!3} \) resonance near the fixed point.
Further outside a \( {10\!:\!3} \) resonance is visible.
Note that the  \( {6\!:\!2} \) resonance chain consists
of two symmetry-related ${3\!:\!1}$ resonance chains which
only differ in the sign of the initial momentum  \( \momx \).
For the billiard system in $\left(\confy, \momy \right)$,
see \prettyref{fig:ps_2d_12}, there is a
\( {3\!:\!1} \) resonance near the fixed point and
further outside a \( {10\!:\!3} \) and a \( {14\!:\!4} \) resonance.
For both \twoD{} billiards, the central regular island is embedded
in a chaotic sea with irregular motion
(blue dots in \prettyref{fig:phasespace2d}).
The central island is
enclosed by a last invariant torus called
boundary circle \cite{GreMacSta1986}.

Note that for both \twoD{}
billiards continuous families of marginally unstable
periodic orbits exist
in the chaotic part of \ps{} at \(
p_i=0 \)~\cite{AltMotKan2005,  TanShu2006, DetGeo2011, BunVel2012}.
Such families are not of relevance for our study,
as they are part of the
recurrence region $\Lambda$ for the \Prs{} discussed in
Sec.~\ref{sec:stickiness}.

Considering the dynamics of the \threeD{} billiard,
beyond the \twoD{} invariant planes, we have to investigate
the full \fourD{} phase space. The central invariant object is
the elliptic-elliptic fixed point
${ \fixedpoint = (\momx, \momy, \confx, \confy) = (0, 0, 0, 0) }$
resulting from a direct product of the fixed points \fixedpointtwod{}
of the two \twoD{} billiards. In the neighborhood of this
elliptic-elliptic fixed point
there is a high density of regular \twoD{} tori \cite{DelGut1996}.
The regular \twotori{} form a whole ``\emph{regular region}'',
similar to the regular islands in the \twoD{} billiards.
However, note that this regular region is
not a connected region but just a
collection of regular tori, permeated by chaotic trajectories
on arbitrarily fine scales,
see Sec.~\ref{sec:channels-AD} for a more detailed discussion.

\subsection{\threeDD{} phase space slice \label{sec:phasespaceslice}}

Since a direct visualization of the \fourD{} \ps{}
of the Poincar\'e section of the three-dimensional billiard is not possible,
we use a \pss{}~\cite{RicLanBaeKet2014} which is defined using a \threeD{}
hyperplane in the \fourD{} \ps{}.
Specifically we choose in the following
\begin{align}
    \label{eq:slice-condition-in-coordinate}
    \Gamma_\varepsilon =
    \left\lbrace (\momx, \momy, \confx, \confy) \; \left|
    \rule{0pt}{2.4ex} \; |\momy| \le \varepsilon \right.
    \right\rbrace
\end{align}
with \( \varepsilon = 10^{-4} \). Whenever a
point $(\momx, \momy, \confx, \confy)$
of an orbit lies within \(\Gamma_\varepsilon\), the remaining
coordinates \((\momx, \confx, \confy)\) are displayed in a \threeD{} plot.
Objects of the \fourD{} \ps{} usually appear in the \pss{}
with a dimension reduced by one. Thus, a typical \twotorus{} leads to
a pair (or more) of \oneD{} lines.
The (numerical) parameter $\varepsilon$ defines
the resolution of the resulting \pss{}.
For smaller $\varepsilon$ longer trajectories
have to be computed to obtain the same number of points
in $\Gamma_\varepsilon$.
For further illustrations and discussions of the \pss{} representation
see Refs.~\cite{RicLanBaeKet2014,LanRicOnkBaeKet2014,OnkLanKetBae2016,AnaBouBae2017}.

\begin{figure}[t]
    \includegraphics{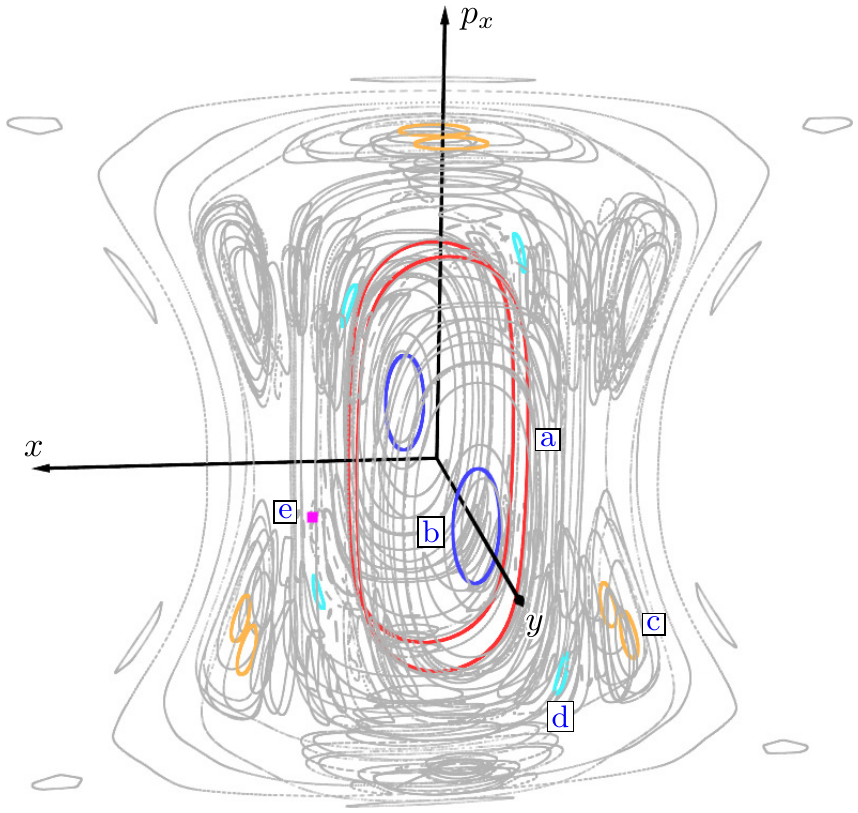}
    \caption{\label{fig:phasespace_focus}
        \pss{} of the billiard
        defined in \prettyref{eq:def_billiard} with $a=1.04$ and
        $b=1.12$
            for
        \( |\momy| \le \varepsilon = 10^{-4} \).
        Regular tori appear as \oneD{} lines (gray).
        For the labeled tori
        (\FBOX{a}--\FBOX{e})
        the corresponding trajectories in \confs{} are shown in
        \prettyref{fig:slice-vs-trajectories}.
     \MOVIEREF{}
    }
\end{figure}

\begin{figure*}[t]
    \subfloat[][]{
        \includegraphics{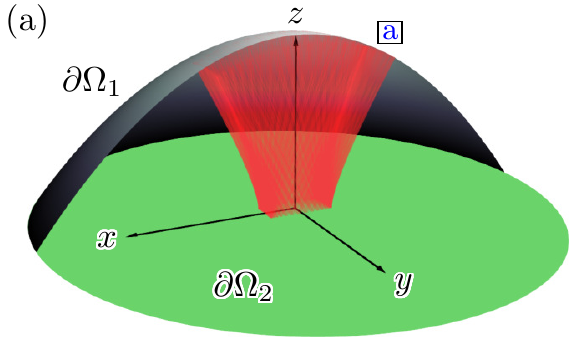}
        \label{fig:regular_torus}
    }
    \subfloat[][]{
        \includegraphics{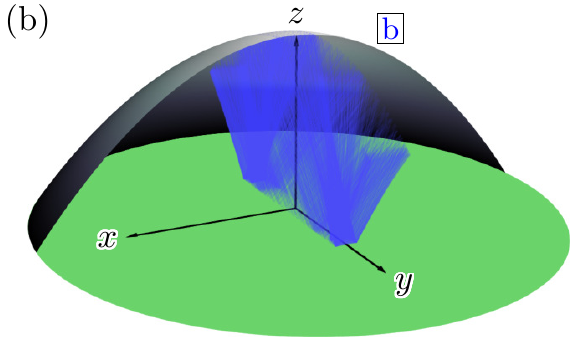}
        \label{fig:regular_torus_out}
    }
    \subfloat[][]{
        \includegraphics{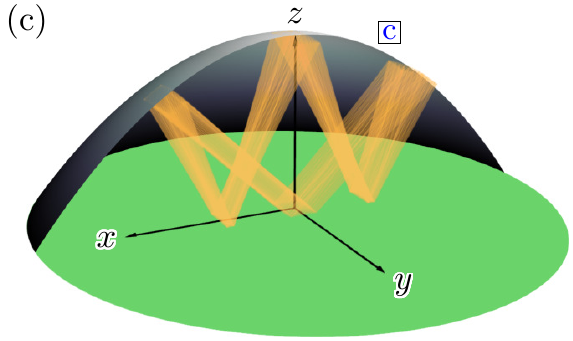}
        \label{fig:regular_island}
    }\\ [-1ex]
    \subfloat[][]{
        \includegraphics{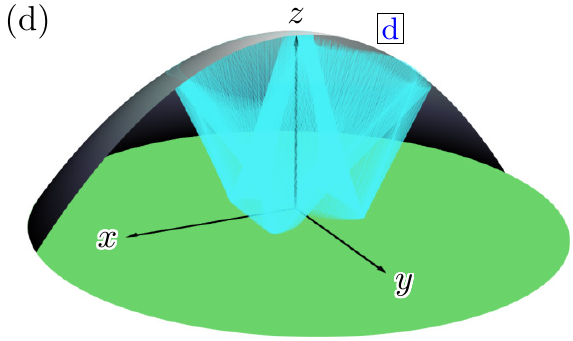}
        \label{fig:regular_curved}
    }
    \subfloat[][]{
        \includegraphics{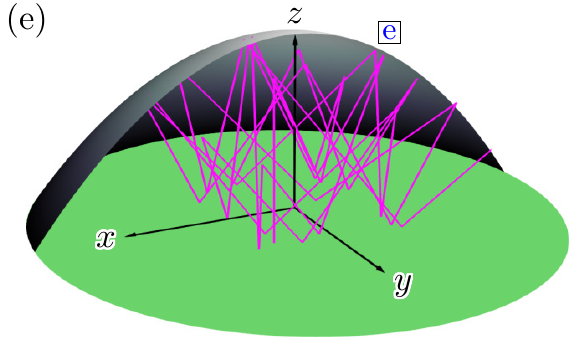}
        \label{fig:double_res}
    }\\
    \caption{\label{fig:slice-vs-trajectories}%
        \threeD{} paraboloid billiard with boundaries $\partial\Omega_2$
        (green) and $\partial\Omega_1$ for which only the part with $y<0$
        and trajectories in configuration space for the
        highlighted tori shown in \prettyref{fig:phasespace_focus}.
        \FBOX{a} Trajectory (red) close to \Mone{}.
        \FBOX{b} Trajectory (blue) close to \Mtwo{}.
        \FBOX{c} Trajectory (orange) of an uncoupled resonance related to
        the billiard system with $a=1.04$.
        \FBOX{d} Trajectory (cyan) of a coupled resonance.
        \FBOX{e} Periodic trajectory (pink) of a double resonance with period
        $35$.
    }
\end{figure*}

\prettyref{fig:phasespace_focus} shows a \pss{} representation
for the \threeD{} paraboloid billiard.
For this a few representative selected
initial conditions on regular tori are iterated until \( 5000 \)
points fulfill the slice condition~\eqref{eq:slice-condition-in-coordinate}.
Regular \twoD{} tori appear as two (or more) distinct rings,
see the colored and labeled tori \FBOX{a}--\FBOX{d}
in \prettyref{fig:phasespace_focus}.
Note that the reflection symmetry of the \threeD{} billiard
at the $x$-$z$~plane leads to a symmetry in the \pss{} with respect to
$x$-$p_x$~plane.
The reflection symmetry at the $y$-$z$~plane
corresponds to a symmetry in the \pss{} with respect to the
$y$-$p_x$~plane.

All regular tori are embedded in a chaotic sea (not shown),
similar to the \twoD{} case in \prettyref{fig:phasespace2d}.
Thus, the chaotic sea is a \fourD{} volume in \ps{} and appears as
a \threeD{} volume in the \pss.

The \pss{} representation resembles in large parts the \twoD{}
phase space $(x, p_x)$ of the \twoD{} billiard
shown in \prettyref{fig:ps_2d_04}. This results from
the chosen slice condition~\eqref{eq:slice-condition-in-coordinate},
$|p_y| \leq \varepsilon$.
Alternatively one could consider the slice condition
$|p_x| \leq \varepsilon$ and display the remaining coordinates $(x, y, p_y)$.
This would resemble in large parts the \twoD{}
phase space $(y, p_y)$ of the \twoD{} billiard
shown in \prettyref{fig:ps_2d_12}.

To obtain a better intuition of the \pss{} we relate some
regular \twoD{} tori of \prettyref{fig:phasespace_focus}
to the corresponding trajectories in configuration space,
see \prettyref{fig:slice-vs-trajectories}:

\FBOX{a}
The pair of red rings correspond to a regular \twoD{} torus
in the \fourD{} phase space and in configuration space to the
trajectory shown in \prettyref{fig:regular_torus}.
This trajectory is close to
the $x$-$z$~plane and may be considered as continuation
of a trajectory of the
\twoD{} billiard dynamics shown in \prettyref{fig:ps_2d_04}
with additional dynamics in $y$-direction.

\FBOX{b} For the \twoD{} torus
shown in \prettyref{fig:regular_torus_out}
the trajectory is close to the $y$-$z$~plane and therefore
similar to the \twoD{} billiard dynamics shown in \prettyref{fig:ps_2d_04}
with additional dynamics in $x$-direction.

\FBOX{c}
The six orange rings correspond to the trajectory of
\prettyref{fig:regular_island}, which is located in the
$3:1$ island chain of \prettyref{fig:ps_2d_04},
again with additional dynamics in $y$-direction.
Note that this trajectory has a symmetry related partner, which is
obtained by inverting the initial momentum.
In the \pss{} this corresponds to the symmetry with respect
to the reflection at the $x$-$y$~plane.

\FBOX{d} A type of dynamics only occurring in \threeD{} billiards is
the cyan \twoD{} torus shown in
\prettyref{fig:phasespace_focus} with trajectory displayed in
\prettyref{fig:regular_curved}. Here the coupling between
both degrees of freedom can be nicely seen in the twisting envelope
of the trajectory in configuration space.
Note that this \twoD{} torus has a symmetry-related partner
obtained by inverting the initial momentum, i.e.\ in
configuration space one obtains the same
type of trajectory passed in opposite sense.
In the \pss{} the symmetry related orbit is obtained
by reflection at the $x$-$p_x$ plane.

\FBOX{e} Moreover, there are also trajectories
of the type shown in \prettyref{fig:double_res}.
This is a periodic orbit with period $35$ extending in
both degrees of freedom and corresponds
to a double resonance (see Sec.~\ref{sec:freqspace}),
which is not possible in a \twoD{} billiard.

The regular \twoD{} tori in the phase space of a \fourD{} map are organized
around families of elliptic
\oneD{} tori~\cite{LanRicOnkBaeKet2014,OnkLanKetBae2016}.
Most prominently one has the so-called
\emph{Lyapunov families} \cite{JorVil1997,JorVil1997b, JorOll2004}
of elliptic \onetori{} which emanate from the central
elliptic-elliptic fixed point \fixedpoint{}.
For the \threeD{} billiard these two families \Mone{} and
\Mtwo{} corresponds to the regular dynamics of the
two embedded \twoD{} billiards shown in \prettyref{fig:phasespace2d}.
These two families of elliptic \oneD{} tori form a ``skeleton''
around which the regular \twotori{} are organized.
For example, the orbit shown in \prettyref{fig:regular_torus}
is a regular \twoD{} torus which is close to the Lyapunov
family \Mone{}, while the orbit in \prettyref{fig:regular_torus_out}
is close to the Lyapunov family \Mtwo{}.

In the chosen \pss{} \eqref{eq:slice-condition-in-coordinate} the
family \Mone{} is completely contained in the $x$-$p_x$~plane of
\prettyref{fig:phasespace_focus}. Note that only a few selected trajectories
of \prettyref{fig:ps_2d_04} are displayed.
In contrast \Mtwo{} coincides with the $y$-axis which is easily seen by
applying the slice condition~\eqref{eq:slice-condition-in-coordinate} to
the phase space shown in \prettyref{fig:ps_2d_12}.
The closeness of the \twoD{} tori shown
in \prettyref{fig:regular_torus} to \Mone{}
and in \prettyref{fig:regular_torus_out}
to \Mtwo{} is also clearly seen
in the \pss{} in \prettyref{fig:phasespace_focus}.
Note that in general the Lyapunov families not necessarily coincide with
conjugate variables of the system, see e.g.~\cite{LanRicOnkBaeKet2014}.

\subsection{Frequency space \label{sec:freqspace}}

\begin{figure*}[!t]
    \includegraphics[width=\textwidth]{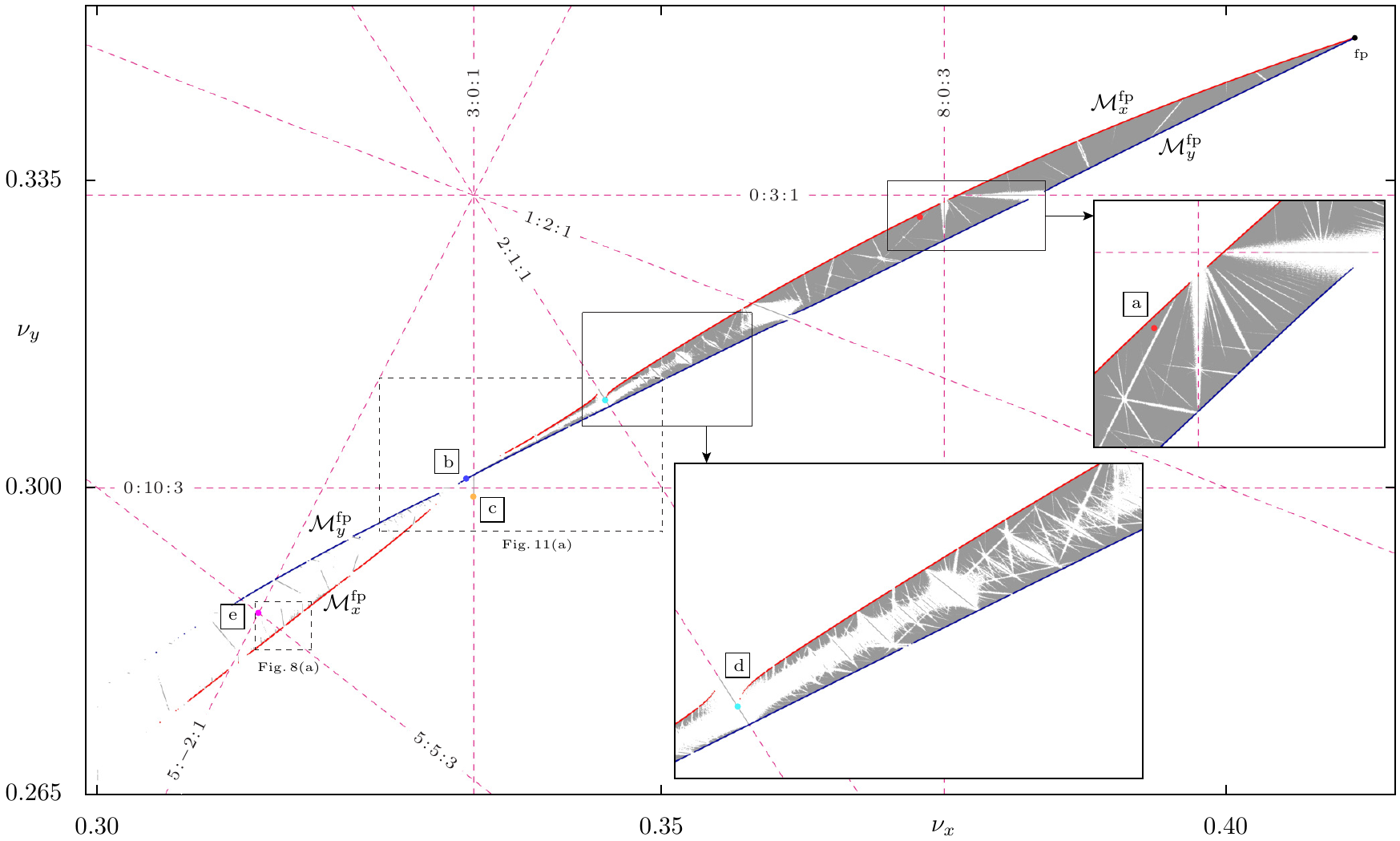}
    \caption{\label{fig:freq_space}%
     Frequency space of the \threeD{} paraboloid billiard defined in
     \prettyref{eq:def_billiard} for $a=1.04$ and $b=1.12$.
     In total $9.6 \times 10^6$ frequencies $(\nuone, \nutwo)$ for
     the \twotori{} are displayed (gray points).
     The rightmost tip \( (\nua, \nub)  = (0.41143, 0.35130) \)
     corresponds to the elliptic-elliptic fixed point \( \fixedpoint \).
     Two Lyapunov families
     of elliptic \onetori{} \Mone{} (red) and \Mtwo{} (blue) emanate from
     this point. Some important resonance lines are shown as magenta dashed lines.
     The insets show magnifications of the frequency space.
     Colored points marked by \FBOX{a}--\FBOX{e}
     correspond to the examples shown in
     \prettyref{fig:phasespace_focus}
     and \prettyref{fig:slice-vs-trajectories}.
        }
\end{figure*}

The frequency space representation is an important complementary
approach for understanding higher-dimensional dynamical systems.
The basic idea is to associate with every
regular \twotorus{} in the \fourD{} phase space its two fundamental frequencies
\( (\nuone, \nutwo ) \in [0, 1)^2 \) and display them in a \twoD{}
frequency space. Numerically this is done using a Fourier-transform based
frequency analysis~\cite{MarDavEzr1987,Las1993, BarBazGioScaTod1996}.
The mapping from phase space to frequency space allows
for explaining features observed in the \pss{}
and identifying resonant motion.

For a given initial condition
an orbit segment $\{\left(\momx^j, \momy^j, \confx^j, \confy^j \right)\}$
with ${j=0, 1, ..., N_{\text{seg}}-1}$
is obtained from consecutive iterates of the map.
In the following $N_{\text{seg}}=4096$ is used.
From this orbit two complex signals
$s_x^j = x^j-\ui p_x^j$ and $s_y^j = y^j-\ui p_y^j$ are constructed,
and for each signal its fundamental frequencies $\nuone$ and $\nutwo$
are calculated.
Note that usually the computed frequencies are
only defined up to an unimodular transformation
\cite{DulMei2003, GekMaiBarUze2007,RicLanBaeKet2014}.
For the considered \threeD{} billiard system
no transformations have to be applied to get a consistent
association in frequency space.

To decide whether the motion for a given initial condition is regular
or chaotic, another orbit segment
$\{\left(\momx^j, \momy^j, \confx^j, \confy^j \right)\}$
for $j=N+N_{\text{seg}},..., N+2N_{\text{seg}}-1$
is computed with $N = 10^5$
giving fundamental frequencies $(\tilde{\nu}_x,  \tilde{\nu}_y)$.
As chaos indicator we use the
frequency criterion
\begin{equation} \label{eq:frequency-criterion}
 \delta = \max{\left(|\nuone - \tilde{\nu}_x|,|\nutwo -
    \tilde{\nu}_y|\right)} < \delta_{\text{reg}}.
\end{equation}
This should be close to zero for a regular orbit,
while for a chaotic orbit the frequencies
of the first and second segment will be very different.
While $N=0$ was used in \cite{RicLanBaeKet2014},
using $N = 10^5$ leads to a more sensitive measure,
in particular excluding short-time transients.
As threshold $\delta_{\text{reg}}=10^{-9}$
has been determined based on a histogram of the $\delta$-values,
computed for many initial conditions together with a visual inspection
of selected orbits in the \pss{}.
This leads to a total of $9.6 \times 10^6$ regular tori
with corresponding frequency pairs shown in \prettyref{fig:freq_space}.
Based on a visual check using the \pss{},
initial points for the sampling of the frequency space are
chosen within an ellipse
$(x, y, z=0)$ with half-axes $r \, a \, \sqrt{2}$, $r \, b \, \sqrt{2}$, and
radius $r=0.8$ (see green region in the inset of \prettyref{fig:recurrence}),
and $p_x^2 + p_y^2 \leq 1$.
Choosing initial conditions outside of this region leads to chaotic
dynamics and thus the frequency criterion \eqref{eq:frequency-criterion}
is not fulfilled.
From the fraction of accepted regular tori
the size of the regular region is estimated as $1.4 \%$ of the
\fourD{} phase space.

Note that even though the frequency criterion \eqref{eq:frequency-criterion}
is a very sensitive chaos-detector, it uses finite-time information
and therefore some of the accepted regular \twoD{} tori
are actually chaotic orbits.
This is of course common to any tool for chaos detection,
see Ref.~\cite{SkoGotLas2016} for a recent overview.

\subsubsection{Regular tori and Lyapunov families}

The geometry of the frequency space is governed by a few organizing elements:\\
First, the frequencies of the central fixed point \fixedpoint{}
can be obtained by a linear stability analysis for each of the
two \twoD{} billiards which gives an analytic expression for the frequency
of \fixedpointtwod{} \cite{SieSte1990}, evaluating to
${\fixedpoint = (\nua, \nub) \approx (0.41143, 0.35130)}$.
This point corresponds to the rightmost tip in the frequency plane
in \prettyref{fig:freq_space}.

Second, the two sharp edges (red and blue)
emanating from the fixed point
correspond to the Lyapunov families of \onetori{} \Mone{} and \Mtwo{}.
For such \onetori{} only the longitudinal frequency \nuL{},
corresponding to \nuone{} for \Mone{} and \nutwo{} for \Mtwo{},
can be determined directly. The
other frequency \nuT, called normal or librating frequency, can be
computed by contracting a surrounding \twotorus{} \cite{LanRicOnkBaeKet2014}
or using a Fourier expansion method \cite{JorOll2004,CasJor2000}.
Going away from the fixed point along the \oneD{} families,
i.e.\ either along \Mone{} or \Mtwo{},
corresponds in \prettyref{fig:phasespace2d} to move from the central
fixed point \fixedpointtwod{} towards the boundary of the regular island.
For the particular geometry of the \threeD{} paraboloid billiard
the Lyapunov families \Mone{} and \Mtwo{}
coincide with the dynamics of the \twoD{} billiards
shown in \prettyref{fig:phasespace2d}.

These lower-dimensional dynamical objects
provide the skeleton of the regular dynamics, both
in frequency space and in phase space,
around which the regular motion on \twoD{} tori is organized.
In the vicinity of the fixed point
\fixedpoint{} the frequency pairs of regular \twoD{} tori
have a high density and quite densely fill the
region between the Lyapunov families, also see upper inset
in \prettyref{fig:freq_space}.
With increasing distance from the fixed point,
e.g.\ below $\nu_x\approx0.35$,
regular \twoD{} tori
only persist close to the families of \onetori{}, see the lower inset
in \prettyref{fig:freq_space}.
Another important observation are the numerous
gaps, i.e.\ regions not covered by regular tori,
which are arranged around straight lines. The origin of these
will be discussed in the following section.

\subsubsection{Resonance lines}

The frequency space is covered by \emph{resonance lines},
on which the frequencies fulfill the resonance condition
\begin{align}
    \label{eq:resonance-driven}
    m_x \nuone + m_y \nutwo = n
\end{align}
for $m_x, m_y, n \in \mathbb{Z}$ without a common divisor and at least
$m_x$ or $m_y$ different from zero.
In the following a resonance condition is denoted as ${m_x:m_y:n}$
and the \emph{order of a resonance} is given by ${|m_x| + |m_y|}$.
The resonance lines form a dense \emph{resonance web}
in frequency space.
Some selected resonance lines are shown in \prettyref{fig:freq_space}.

For a given frequency pair $(\nuone, \nutwo)$
the number of independent resonance conditions, the so-called rank,
determines the type of motion occurring in
\fourD{} maps~\cite{Tod1994,Tod1996} (also see \cite{LanRicOnkBaeKet2014}
for an illustration):

\begin{itemize}
  \item If the frequency pair fulfills no resonance condition,
        it is of rank-0 and the motion on the corresponding \twoD{}
        torus is quasi-periodic, filling it densely.
        Such frequencies for example correspond to KAM tori
        of sufficiently incommensurate frequencies.
        Examples are the red (\FBOX{a}) and blue (\FBOX{b})
        marked points which correspond to the tori in \pss{} shown in
        \prettyref{fig:phasespace_focus} and the trajectories in
        \prettyref{fig:regular_torus} and \prettyref{fig:regular_torus_out}.

  \item If only one resonance condition is fullfilled (\mbox{rank-1} case),
        the resonance is either
        (a) uncoupled, i.e.\ ${m_x\!:\!0\!:\!n}$ or ${0:m_y:n}$, or
        (b) coupled, i.e.\ ${m_x\!:\!m_y\!:\!n}$ with
        both $m_x$ and $m_y$ non-zero.
        The motion is quasi-periodic
        on a \oneD{} invariant set
        which either consists of one
        component in the case of coupled resonances,
        or of $m_x$ (or $m_y$) dynamically connected
        components in the case of uncoupled resonances.
        Note that in this rank-1 case one has (at least)
        one pair of elliptic and hyperbolic \oneD{} tori
        \cite{OnkLanKetBae2016}.

        An example of an uncoupled resonance is the
        orange marked frequency pair
        $(\nuone, \nutwo) = (0.3333, 0.2989)$
        located on the \( {3\!:\!0\!:\!1} \) resonance line,
        corresponding to the orange torus of \prettyref{fig:phasespace_focus}
        and the trajectory in \prettyref{fig:regular_island}.

        An example of a coupled resonance is
        the cyan frequency pair $(\nuone, \nutwo) = (0.3450, 0.3099)$
        located on the ${2\!:\!1\!:\!1}$ resonance line,
        corresponding to the cyan
        colored torus in \prettyref{fig:phasespace_focus}
        and the trajectory in \prettyref{fig:regular_curved}.

        The difference between uncoupled and coupled
        resonances can also be seen by
        \threeD{} projections encoding the value of the projected
        coordinate by color scale \cite{PatZac1994,KatPat2011},
        see e.g.~Fig.~5 in~\cite{LanRicOnkBaeKet2014} for a detailed
        illustration.

  \item If two independent resonance conditions are fulfilled (rank-2 case)
        one has a double resonance.
        The frequency pair lies at the intersection
        of two resonance lines
        and leads to (at least) four periodic orbits with
        different possibilities for their stability.

        As an example, we consider the frequency pair
        $(\nuone, \nutwo) = (\nicefrac{n_1}{m_1}, \nicefrac{n_2}{m_2} )
                          = (\nicefrac{11}{35}, \nicefrac{2}{7} ) $
       in \prettyref{fig:freq_space}
       which is the intersection of the
       ${5\!:\!5\!:\!3}$ and ${5\!:\!-2\!:\!1}$ resonance line.
       The period is given by $\text{lcm}(m_1, m_2) = 35$.
       The corresponding elliptic-elliptic trajectory is shown
       in configuration space in
       \prettyref{fig:double_res}.

\end{itemize}

Resonances also lead to gaps within the areas covered
by regular \twoD{} tori, see e.g.\
the white regions in \prettyref{fig:freq_space}.
Of particular strong influence are resonances of low order
as they typically lead to the largest gaps.
When resonance lines intersect the families of \oneD{} tori,
this also leads to gaps within these families.
So strictly speaking, they form one-parameter Cantor families
of \oneD{} tori \cite{JorVil1997,JorVil2001}.
If either $|m_x|\le 2$ (for \Mone) or $|m_y|\le 2$ (for \Mtwo)
this leads to gaps or strong bends in the corresponding families of \oneD{}
tori~\cite{OnkLanKetBae2016}.
For example,
due to the \( {3\!:\!0\!:\!1} \) resonance there is a large gap in \Mone{} and
due to the \( {0\!:\!10\!:\!3} \) resonance a smaller one in \Mtwo{}.
Note that in the case of the considered \threeD{} paraboloid billiard
the frequencies of the families of \oneD{} tori cross
near these gaps, see \prettyref{fig:freq_space}.

\subsection{Hierarchy}\label{sec:hierarchyps}

For systems with two degrees of freedom
the phase space shows a hierarchy of
islands-around-islands on ever finer scale
which are organized around elliptic periodic orbits \cite{Mei1986}.
In higher-dimensional systems the organization
of phase space is based on higher-dimensional elliptic objects.
For example for a \fourD{} map, families of elliptic
\oneD{} tori form the skeleton of surrounding regular \twoD{} tori,
as discussed above.
Thus the generalization of the island-around-island
hierarchy can be fully described in terms
of the families of elliptic \oneD{} tori \cite{LanRicOnkBaeKet2014}:
There are two possible origins of such families which either
$\left(\alpha \right)$ emanate from an elliptic-elliptic periodic point or
$\left(\beta \right)$ result from a family of broken \twotori{}
fulfilling a rank-1 resonance condition.
For the first case one further distinguishes:
$\left(\alpha 1 \right)$ the fixed point is either the central
elliptic-elliptic fixed point \fixedpoint{} or it corresponds to an
elliptic-elliptic periodic point resulting from a broken \twotorus{}
which fulfills a rank-2 resonance.
$\left(\alpha 2 \right)$
The families of \oneD{} tori emerge from
an elliptic-elliptic periodic point resulting
from a broken elliptic \onetorus{} when its
longitudinal frequency $\nuL = \frac{n}{m}$
fulfills an rank-1 resonance.
This corresponds to an intersection of
a resonance line with a one-parameter family of elliptic \oneD{} tori.

As this hierarchy of elliptic \oneD{} tori
is reflected in the surrounding \twoD{} tori,
Figs.~\ref{fig:phasespace_focus}--\ref{fig:freq_space}
provide an illustration of the hierarchy:
\begin{itemize}
\item $\left(\alpha 1 \right)$:
     \FBOX{a} and \FBOX{b} are examples for orbits close to
     \Mone{} and \Mtwo{}, respectively, which emanate from the central
     elliptic-elliptic fixed point \fixedpoint{}.
     An example of a double resonance is
     the elliptic-elliptic periodic point shown in \FBOX{e}.
     From this periodic orbit also
     two Lyapunov families of elliptic \oneD{} tori emerge.
\item $\left(\alpha 2 \right)$:
    \FBOX{c} is an example in the surrounding of the period-3 island,
    where one elliptic \oneD{} torus of \Mone{} with
    longitudinal frequency $\nuL = \frac{1}{3}$
    fulfills the ${3\!:\!0\!:\!1}$ resonance (rank-1), which gives rise
    to a period-3 periodic orbit with attached Lyapunov families.
\item $\left(\beta \right)$:
    An example of a two-parameter family of \twotori{}
    fulfilling a rank-1 resonance condition is the ${2\!:\!1\!:\!1}$ resonance,
    for which \FBOX{d} shows one surrounding \twotorus{}.
\end{itemize}
Analyzing the dynamics of this hierarchy
in frequency space requires an adjusted frequency analysis as the
frequencies collapse to either $(\alpha)$ a point or $(\beta)$
a resonance line \cite{LanRicOnkBaeKet2014}.

\subsection{Resonance channels and Arnold diffusion}
\label{sec:channels-AD}

Points on a resonance line
correspond in phase space either to elliptic \oneD{} tori
or the surrounding \twoD{} tori.
Thus the one-parameter family of elliptic \oneD{} tori
forms the ``skeleton'' of the so-called \emph{resonance channel}.
The regular part of the resonance channel
consists of the  elliptic \oneD{} tori and their surrounding \twoD{} tori.
The chaotic part of the resonance channel
consists of the corresponding hyperbolic \oneD{} tori
and the chaotic motion in the \emph{stochastic layer},
which is associated with the homoclinic tangle
of the stable and unstable manifolds
of the hyperbolic \oneD{}~tori.
For a detailed discussion
of the geometry of resonance channels in phase space
and the relation to bifurcations of families
of elliptic \oneD{} tori see \cite{OnkLanKetBae2016}.

In these stochastic layers chaotic transport
along the resonance channels is possible,
which is commonly referred to as \emph{Arnold diffusion}
\cite{Arn1964,Chi1979,Loc1999,Dum2014}.
As resonance lines cover the whole frequency space
densely, all stochastic regions of phase space are connected.
Their network within the region of regular tori is referred to
as \emph{Arnold web}.

Perturbing an integrable system,
Nekhoroshev theory shows that in the near-integrable regime
the speed of Arnold diffusion is exponentially small \cite{Nek1977,Guz2004},
which makes its numerical detection very difficult.
This regime is called \emph{Nekhoroshev regime}.
For stronger perturbations regular tori become sparse
and neighboring stochastic layers begin to overlap
with much faster transport,
in particular across channels. This regime
is called \emph{Chirikov regime} \cite{GuzLegFro2002,FroLegGuz2006}.

The considered \threeD{} paraboloid billard does not qualify as
near-integrable.
Still the dynamics within a given resonance channel
shows both the behavior of the Nekhoroshev and the Chirikov regime
depending on the location along the channel,
see e.g.\ lower inset of \prettyref{fig:freq_space}:
Near the intersection of the resonance line
with \Mone{} or \Mtwo{},
the stochastic layer is embedded in surrounding
regular \twoD{} tori and the chaotic dynamics along the
channel should be governed by the slow Arnold diffusion, i.e.\
this part of the resonance channel belongs to the Nekhoroshev regime.
Further along the channel
the neighboring regular tori become more sparse
and one gets into the Chirikov regime
in which the stochastic layers of neighboring resonances
overlap.
Thus transport across resonance channels becomes possible
and is more likely further along the channel.
In addition other crossing resonance channels may also be explored
by a trajectory started within a stochastic layer.

With this general background in mind it is also
possible to represent chaotic trajectories in frequency space
and interpret the results:
Of course, for chaotic trajectories in a stochastic layer no frequencies
exist in the infinite time limit,
however it is possible to associate ``finite-time''
frequencies.
For example chaotic trajectories approaching a regular torus
also acquire similar frequencies.
And for trajectories in a stochastic layer their
finite-time frequencies will cover a small region
in the surrounding of the corresponding resonance line in frequency space.
This will be illustrated and discussed in detail in
Sec.~\ref{sec:long_trapped}.

\section{Stickiness and power-law trapping} \label{sec:stickiness}

\subsection{Poincar\'e recurrence statistics} \label{sec:prs}

In systems with a mixed phase space the transport between
different regions can be strongly slowed down by so-called
\emph{stickiness} of chaotic trajectories taking place
in the surrounding of regular regions.
A convenient approach to characterize stickiness is based on the \emph{\Prt}.
It states that for a measure-preserving map
with invariant probability measure $\mu$ almost all
orbits started in a region \initregion{} of phase space will return to
that region at some later time~\cite{Poi1890}.
Based on the recurrence times $t_{\text{rec}}(x)$
of orbits with initial conditions $x\in\initregion$,
one obtains the \emph{recurrence time distribution}
\begin{equation} \label{eq:rtd}
  \rho(t) = \frac{\mu\left( x \in \initregion \; | \;
                            t_{\text{rec}}(x) = t \right)}
                 {\mu(\initregion)}.
\end{equation}
The average recurrence time follows from
Kac's lemma \cite{Kac1947,Kac1959,Mei1997} as
\begin{equation} \label{eq:Kac}
  \langle t_{\text{rec}} \rangle
    := \frac{1}{\mu(\initregion)} \int\limits_{\Lambda}
           t_{\text{rec}}(x) \; \ud \mu
     = \frac{\mu(M_{\text{acc}})}{\mu(\initregion)},
\end{equation}
where $M_{\text{acc}}$ is the accessible region
for orbits starting in region $\initregion$.

Instead of considering the distribution of recurrence times,
it is numerically more convenient to use the
\emph{Poincar\'e recurrence statistics},
which is the complementary cumulative Poincar\'e recurrence time
distribution,
\begin{equation}
   P(t)  = \sum_{k=t}^\infty \rho(k),
\end{equation}
i.e.\ the distribution of the recurrence times larger than $t$.
Initially one has $P(0)=1$ and by definition
$P(t)$ is monotonically decreasing.
Numerically, the \Prs{} is determined by
\[
   P\left( t \right)  = \frac{N \left(t\right)}{N(0)} ,
\]
where \(N(0)\) is the number of trajectories initially started in
\initregion{} and \( N\left(t\right) \) is the number of trajectories
which have not yet returned to \initregion{} until time $t$.

The nature of the decay of $P(t)$ depends on the
dynamical properties of the systems. Fully chaotic systems show an exponential
decay \cite{BauBer1990, ZasTip1991,HirSauVai1999, AltSilCal2004}
whereas generic systems with a mixed phase space
typically exhibit a \pol{} decay~\cite{
ChaLeb1980,ChiShe1983, Kar1983, ChiShe1984, KayMeiPer1984a, KayMeiPer1984b,
HanCarMei1985,
MeiOtt1985, MeiOtt1986, ZasEdeNiy1997, BenKasWhiZas1997, ChiShe1999,
ZasEde2000, WeiHufKet2002, WeiHufKet2003, CriKet2008,
CedAga2013, AluFisMei2014,AluFis2015, AluFisMei2017}.
Note that considering the Poincar\'e recurrence statistics
with respect to  \initregion{}
can also be seen as an escape experiment from an open billiard
so that the decay of $P(t)$ agrees with the decay
of the survival probability with trajectories injected
in the opening \initregion{} \cite{AltTel2008}.

To numerically study the \Prs{} the  region
\initregion{} in phase space should fulfill two prerequisites in order
to obtain good statistics:
First
\initregion{} should be placed in the chaotic sea far away from the regular
region to ensure that trajectories are started outside of the expected
sticky region. Second, the volume of \initregion{} should be chosen
sufficiently large to ensure that non-trapped orbits return quickly enough
to avoid unnecessary computations.
For the \fourD{} Poincar\'e map of the \threeD{} billiard we chose
\begin{align}\label{eq:intialregion}
    \begin{split}
\Lambda = &
            \bigg\{
            \left( \momx, \momy, \confx, \confy \right): \:
             \frac{1}{2} \left( \left(\frac{\confx}{a}\right)^2
             + \left(\frac{\confy}{b}\right)^2\right) > r^2 ,\\
          &
          ~~ \left( x, y \right) \in \partial \Omega_2, \text{ and }
                p_x^2 + p_y^2 \le 1
               \bigg\}.
    \end{split}
\end{align}
A point $(p_x, p_y, x, y) \in \Lambda$ defines the initial condition
$(p_x, p_y, p_z, x, y, z)$ with $z=0$ and $p_z > 0$ via $||\mathbf{p}||=1$.
In configuration space the region $\Lambda$ corresponds to an elliptical ring
in the $z=0$ plane, defining the opening in \ground{},
marked in yellow in the inset of \prettyref{fig:recurrence}.
This allows for starting trajectories into the billiard under
all different angles.
After visual inspection of the regular structures with the help of the
\pss{} we choose \( r=0.8 \) for
\initregion{}, which covers \(36 \%\) of the \fourD{} phase space of
the \poincare{} map.

For the determination of the \Prs{}
\( {N(0) = 10^{13} }\) random initial conditions are chosen
uniformly in $\Lambda$. For each of them the real flight time and
the number of iterations of the Poincar\'e map
are determined until the trajectory returns to
\initregion{}.

In \prettyref{fig:recurrence} the Poincar\'e recurrence
statistics for real flight times and the number of mappings is shown.
Initially one has approximately an exponential decay for small
times up to \({ t \lesssim 200 } \).
This corresponds
to chaotic trajectories which are not trapped near
any of the regular structures and thus return to $\Lambda$
very quickly.
For larger times $P(t)$ exhibits an overall \pol{} decay
${P(t) \sim t^{-\gamma}}$
with exponent  \( \gamma \approx 1.2 \).
The only exception of the straight \pol{}
is a small step for  \( t \in [ 10^{6}, 10^{7}] \)
which could be a manifestation of some more restrictive
partial barriers.
Note that one could also consider
other geometries of the opening which only affects
the initial exponential decay but not the exponent of the \pol{}
decay \cite{Fir2014}.
The \Prs{} of the number of mappings $t$ and real flight time $\tau$
are shifted by
approximately a factor of $\sim 1.94$
which is close to the geometric length $\tau=2$ of the
stable periodic orbit in the center of the billiard.

An interesting application of Kac's lemma is
to estimate the size $\mu(M_{\text{reg}})$ of
the regular region $M_{\text{reg}}$, as it was done
in \cite{Mei1997} for the \twoD{} H\'enon map.
Using the average recurrence time $\langle t_{\text{rec}} \rangle$
one gets from \eqref{eq:Kac}
\begin{equation} \label{eq:M-reg-via-Kac}
   \mu(M_{\text{reg}}) = 1 - \mu(M_{\text{acc}})
                     = 1 - \langle t_{\text{rec}} \rangle  \mu(\Lambda).
\end{equation}
With $\langle t_{\text{rec}} \rangle = 2.696$
and $\mu(\Lambda) = 0.36$ we obtain
$\mu(M_{\text{reg}}) = 0.029$
which gives an estimate of $2.9\%$ for the size of the regular region
in the \fourD{} phase space.
This is approximately twice as large as the regular fraction
determined using the frequency criterion, see \prettyref{sec:freqspace}.
Note that using \prettyref{eq:M-reg-via-Kac}
is expected to provide an upper bound to the size
of the regular region, as orbits started in \initregion{}
explore the chaotic region and thus approach the regular
region from the outside.
Moreover, there might be chaotic regions which are not accessed
at all on the considered time-scales,
while initial points in such regions could
already be detected as chaotic by the frequency criterion
\eqref{eq:frequency-criterion}.
Moreover, the  threshold $\delta_{\text{reg}}$ for the
frequency criterion has been chosen quite small
and relaxing this to $\delta_{\text{reg}}=10^{-7}$ gives
comparable results for the size of the regular region.

The overall slow decay in the \Prs{} is due to orbits
with large recurrence times $t_{\text{rec}}$. Therefore we want to
analyze such long-trapped orbits within the phase space and frequency space
introduced in \prettyref{sec:phasespaceslice} and \prettyref{sec:freqspace}.

\subsection{Long-trapped orbits}\label{sec:long_trapped}

To obtain a better understanding of
the origin of the observed power-law decay of the \Prs{} we
consider one representative example of a
long-trapped chaotic orbit in the following.
Analyzing this long-trapped orbit both in the \pss{}
and in frequency space
allows to draw the following conclusions:
Trapping takes place
(i) at the ''surface'' of the regular region (outside the Arnold web)
and is (ii) not due to a generalized island-around-island hierarchy,
as discussed in Sec.~\ref{sec:hierarchyps}.
We find that the dynamics of long-trapped orbits is (iii)
governed by numerous resonance channels which extend far into the chaotic sea.
The results suggest to decompose the dynamics in the
sticky region into (iii.a) transport across resonance channels
and (iii.b) transport along resonance channels.
For the transport across resonance we find clear
signatures of partial barriers.
All these points support the results obtained in Ref.~\cite{LanBaeKet2016}
for the case of the \fourD{} coupled standard map.
In particular we obtain a very clear example
of the geometry of the trapped orbit in the \pss{}
and its signature in frequency space.
Note that we only show one representative orbit here,
while the essential features are observed for all of the about 100 orbits we
analyzed.

\begin{figure}[t]
    \includegraphics{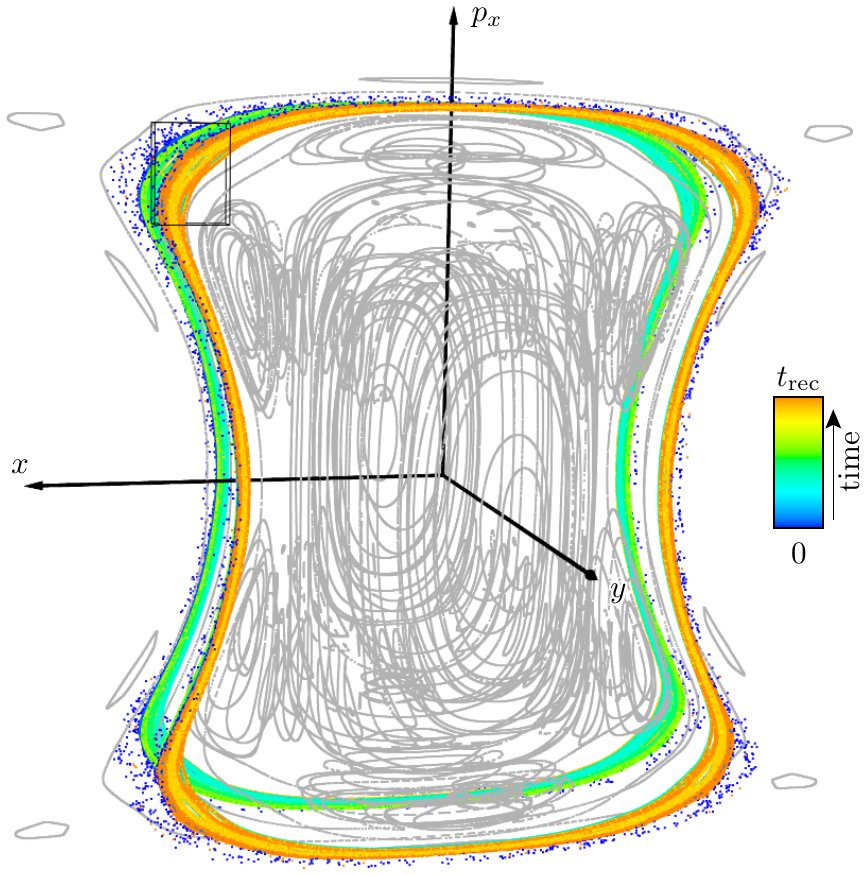}
    \caption{\label{fig:trapped_ps_one}
        Trapped orbit with recurrence time
        $t_{\text{rec}}\simeq1.6 \times 10^9$
        in the \pss{} representation with
        time encoded in color.
        The box indicates the magnification shown in
        \prettyref{fig:trapped_orbit_one_extent}.
      \MOVIEREF{}
    }
\end{figure}

\begin{figure*}[t]
    \includegraphics{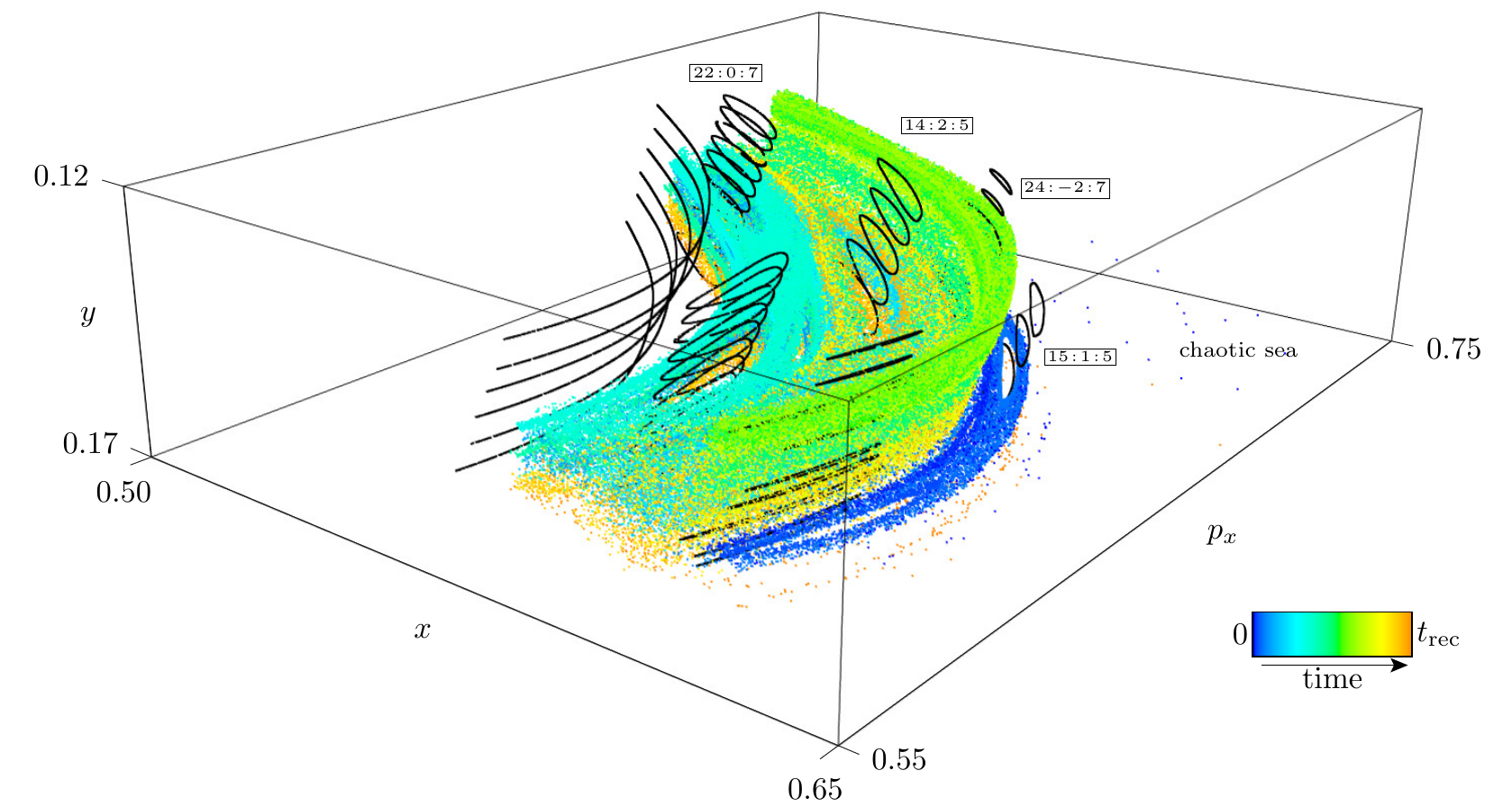}
    \caption{\label{fig:trapped_orbit_one_extent}
        Magnification (rotated) of the trapped orbit of
        \prettyref{fig:trapped_ps_one}
        with recurrence time $t_{\text{rec}}\simeq1.6 \times 10^9$
        in the \pss{} with
        time $[ 0, t_{\text{rec}} )$ encoded in color.
        Regular tori of some important
        resonance channels ${22\!:\!0\!:\!7}$, ${14\!:\!2\!:\!5}$,
        ${24\!:\!-2\!:\!7}$ and ${15\!:\!1\!:\!5}$
        are shown in black.
        The sticky orbit approaches the regular structures
        by going across several resonance channels.
        For the time-resolved animation see
        \MOVIELINK{}.
    }
\end{figure*}

To arrive at these conclusions we make use of two
time-resolved representations of the long-trapped orbit.
In the \pss{} points of the long-trapped orbit are colored according to time
(from blue at $t=0$, starting in $\Lambda$, to orange at $t=t_{\text{rec}}$,
returning to $\Lambda$, see colorbar in \prettyref{fig:trapped_ps_one}).
It is also possible to analyze trapped orbits
in frequency
space~\cite{MarDavEzr1987,Las1993,ConHarVog2000,Lan2016,LanBaeKet2016}.
Although for chaotic
orbits no fundamental frequencies exist, a numerical assignment
of frequencies is still possible because for short time intervals
the dynamics of nearby regular tori is resembled.
For this a sticky chaotic orbit is divided into segments
of length \( N_{\text{seg}} = 4096 \).
For each segment the frequency analysis, see
\prettyref{sec:freqspace}, is performed.
This leads to a sequence of
consecutive frequencies
\( ( \nuone(t_i), \nutwo(t_i)) \)
with $t_i = i N_{\text{seg}}$, $i\in\N$,
which can be displayed either in frequency space with time encoded in color
or as frequency-time signals,
see \prettyref{fig:trapped_orbit_one_freq_space-all}.
Exemplarily, we consider one
long-trapped chaotic orbit with large recurrence time
$t_{\text{rec}}\simeq1.6 \times 10^9$ in the \pss{},
see \prettyref{fig:trapped_ps_one}
and \prettyref{fig:trapped_orbit_one_extent},
and in frequency space, see \prettyref{fig:trapped_orbit_one_freq_space-all};
another example is discussed in App.~\ref{app:trapped-2},
\prettyref{fig:trapped_orbit_two_freq_space}.
Note that these trapped orbits are shown in a
\pss{} with slice parameter $\varepsilon = 10^{-3}$, see
\prettyref{eq:slice-condition-in-coordinate}, to obtain a higher
density of points.

\subsubsection*{(i) Trapping is at the surface of the regular region}

The long-trapped orbit is shown in the time-encoded \pss{} in
\prettyref{fig:trapped_ps_one}.
It is close to the $x$-$p_x$~plane and thus close to \Mone{}.
The coloring of the orbit according to time shows
several bands with different colors, which means that the long-trapped orbit
covers different regions of phase space for specific time intervals.
Furthermore it is close to regular
phase-space structures (gray).
More precisely, the orbit is located close to the ``surface''
of the regular region, which is composed of the regular \twoD{} tori
shown as grey rings in the \pss.
This is even better seen in a magnified (and rotated) sideways view
of the box indicated in the upper
left part in \prettyref{fig:trapped_ps_one}.
This magnification is shown in \prettyref{fig:trapped_orbit_one_extent},
where the surface of the regular region is indicated
by the regular \twoD{} tori (black lines) at the left side.
Going towards the regular region corresponds to decreasing $x$
and $p_x$.
The long-trapped orbit arrives
from the chaotic see, i.e.\ from the right in the figure,
and approaches the regular region while filling
several bands before it returns to the initial region $\Lambda$.
An animation of the time-evolution of the long-trapped orbit
is provided in the Supplemental Material at \MOVIELINK.
Further conclusions which can be drawn from this magnification
will be discussed below.

\begin{figure*}
    \includegraphics{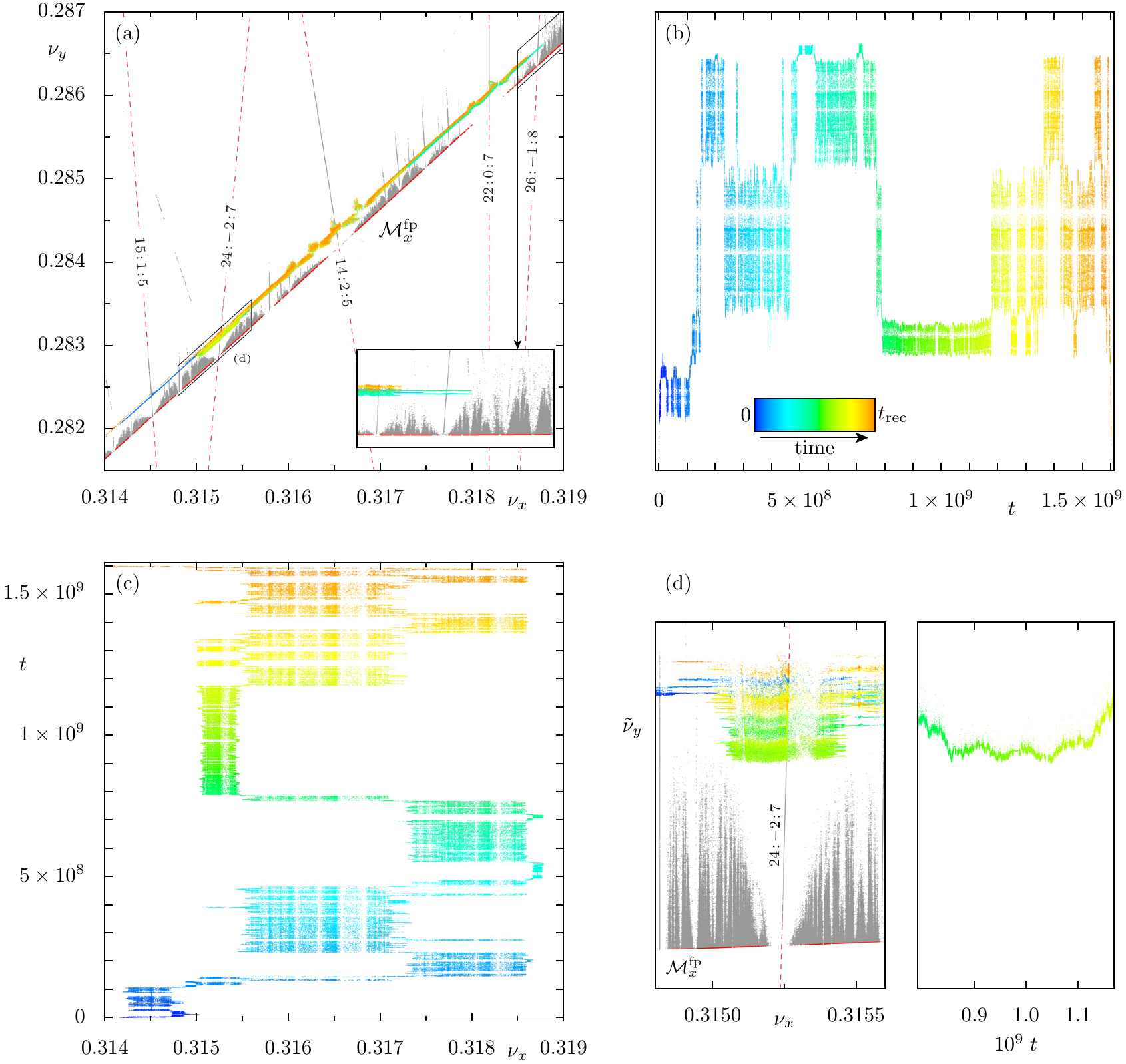}
    \caption{\label{fig:trapped_orbit_one_freq_space-all}
        Frequency space representation of the trapped orbit,
        see \prettyref{fig:trapped_orbit_one_extent}, with
        $t_{\text{rec}} = 1.6 \times 10^{9}$  and time encoded in color.
        (a) Magnification of \prettyref{fig:freq_space}
        with regular tori (grey dots) and selected resonance lines,
        (b, c) frequency-time signals \( \nutwo(t) \) and
        \( \nuone(t) \), respectively,
        (d) magnification of frequency-space, see box in (a),
        in local coordinates.
    }
\end{figure*}

In frequency space it can also be seen that the long-trapped orbit is
located close to the surface of the regular region:
Here the segment-wise determined frequencies of the orbit
extend over a large region,
see \prettyref{fig:trapped_orbit_one_freq_space-all}(a),
which shows a magnification of the frequency space
in \prettyref{fig:freq_space}
with the frequencies of the trapped orbit colored according to time.
Moreover, $9 \times 10^6$ additional frequency pairs
of regular \twoD{} tori are shown (grey dots).
The long-trapped orbit spreads approximately parallel to the
Lyapunov family \Mone{}, staying above the associated regular tori.
\prettyref{fig:trapped_orbit_one_freq_space-all}(d)
shows a magnification of the region indicated in
\prettyref{fig:trapped_orbit_one_freq_space-all}(a),
where the ordinate $\tilde{\nu}_y$ is the distance to the
lower side of the parallelogram, $\tilde{\nu}_y = \nu_y + k\cdot \nu_x + \nu_s$
with $k=-0.99$ and $\nu_s = 0.02925$.
Recall that the lower edge (red), which corresponds to
the family of \oneD{} tori \Mone{},
can be considered as inner part of the regular region.
Thus decreasing $\tilde{\nu}_y$ moves towards the surface
of the regular region.
Moreover, increasing $\nuone$
moves towards the central elliptic-elliptic fixed point.
As the sticky orbit stays well outside of
any regions with many \twoD{} tori,
it is effectively trapped at the surface of the regular
region.

In particular this means that
it does not enter the Arnold web of
resonance lines
which are embedded within regular tori.

\subsubsection*{(ii) Trapping is not due to a hierarchy}

Trapping is also not due to the generalized
island-around-island hierarchy, summarized in Sec.~\ref{sec:hierarchyps}.
This can be concluded from the \pss{} representation. Trapping
deep in a hierarchy would imply successive scaling on finer and finer
phase-space structures as known from \twoD{} maps \cite{MeiOtt1985}.
However, the long-trapped orbit spreads
over the surface of the regular region and no signatures of a
hierarchy are visible.
This is also supported by the frequency-time signals
$\nuone(t)$ and $\nutwo(t)$ shown in
\prettyref{fig:trapped_orbit_one_freq_space-all}(b,~c).
For trapping in the generalized hierarchy the frequencies either collapse on a
frequency pair $(\alpha_1)$ or on a resonance line $(\alpha_2, \beta)$
\cite{LanRicOnkBaeKet2014}.
Neither of these is observed for the considered example.
Note that for the second example of a long-trapped orbit
discussed in App.~\ref{app:trapped-2},
the frequency-time signals shown in
\prettyref{fig:trapped_orbit_two_freq_space}(b,~c)
collapse on the ${3\!:\!0\!:\!1}$ resonance for some longer time interval.
Still the trapping is not dominated by a hierarchy.

\subsubsection*{(iii) Resonance channels}

The frequency-time signals shown in
\prettyref{fig:trapped_orbit_one_freq_space-all}(b,~c) do not collapse
on a specific frequency or resonance line
but mainly fluctuate within specific frequency ranges
over longer time intervals.
These frequency ranges are confined around certain resonance lines,
as shown in \prettyref{fig:trapped_orbit_one_freq_space-all}(a),
for which ${15\!:\!1\!:\!15}$, ${24\!:\!-2\!:\!7}$,
${14\!:\!2\!:\!5}$, ${22\!:\!0\!:\!7}$, and
${26\!:\!-1\!:\!8}$ are the most important resonances.

These resonance lines correspond to regular dynamics,
as discussed in Sec.~\ref{sec:channels-AD}.
Some selected examples of corresponding regular \twoD{} tori
are displayed as stacks of black rings in the \pss{} representation
in \prettyref{fig:trapped_orbit_one_extent}.
For the long-trapped orbit the bands of similarly colored
points in the \pss{} are arranged around
these regular parts of the resonance channels.
This suggests that the long-trapped orbit is confined
to the stochastic layer of the resonance channels.

Both representations and in particular the transformation
to local coordinates $(\nuone, \tilde{\nu}_y)$ as shown
in \prettyref{fig:trapped_orbit_one_freq_space-all}(d),
together with the animation
of the long-trapped orbit in \prettyref{fig:trapped_orbit_one_extent},
suggest a decomposition
of the chaotic dynamics in transport (iii.a) across and (iii.b) along resonance
channels:

(iii.a) In the \pss{} of \prettyref{fig:trapped_orbit_one_extent}
the distinct colored bands indicate that the sticky
orbit stays for extended time intervals within the stochastic layer of a given
resonance channel, e.g.\ ${14\!:\!2\!:\!5}$ or ${22\!:\!0\!:\!7}$,
and then quickly jumps to a different resonance channel.
This happens mainly within the $x$-$p_x$~plane, i.e.\ for approximately
constant $y$.
These transitions across different resonance channels
are also clearly seen in frequency space in
\prettyref{fig:trapped_orbit_one_freq_space-all}(b,~c).

(iii.b)
Transport along resonance channels is best seen
for the ${24\!:\!-2\!:\!7}$ resonance, see
the green colored band in \prettyref{fig:trapped_orbit_one_extent},
which extends in $y$-direction.
In frequency space this corresponds to the magnification
shown in \prettyref{fig:trapped_orbit_one_freq_space-all}(d).
We now discuss both types of transport in more detail.

\subsubsection*{(iii.a) Across resonance channels}
In \prettyref{fig:trapped_orbit_one_freq_space-all}(a) and in particular
in the frequency-time signals
 $\nuone(t)$ and $\nutwo(t)$ the importance of four resonance lines
namely ${15\!:\!1\!:\!5}$, ${24\!:\!-2\!:\!7}$,
${14\!:\!2\!:\!5}$ and ${22\!:\!0\!:\!7}$ are clearly visible.
These resonance lines, together with the surrounding
stochastic layers, form the resonance channels which
define frequency intervals that are covered in a random looking
manner. Furthermore fast
transitions to other frequency intervals are observed.
The long-trapped orbit is initially, up to $t\approx 10^8$,
mainly in an interval around the ${15\!:\!1\!:\!5}$ resonance
and then up to $t\approx 2.5 \times 10^8$ around
the ${22\!:\!0\!:\!7}$ resonance, followed by a longer time window
up to $t\approx 5 \times 10^8$ around the ${14\!:\!2\!:\!5}$ resonance.
Subsequent frequency intervals are around the ${24\!:\!-2\!:\!7}$,
${14\!:\!2\!:\!5}$, and ${22\!:\!0\!:\!7}$ resonance,
sometimes with short excursions to the other intervals.
The closest approach to the regular region
corresponds to the rightmost tip in
\prettyref{fig:trapped_orbit_one_freq_space-all}(a),
i.e.\ largest values of \nuone{} and \nutwo{} in
\prettyref{fig:trapped_orbit_one_freq_space-all}(b,~c).
For short time intervals (cyan and green)
the small stochastic layer around the ${26\!:\!-1\!:\!8}$
resonance is accessed.
Going further to the right is effectively blocked by
a region containing many regular \twoD{} tori as
indicated in the inset of
\prettyref{fig:trapped_orbit_one_freq_space-all}(a).
Even though this region is threaded by resonance
lines on arbitrarily fine scales, the effective transport
along these lines is expected to be very slow.
This is also suggested by the geometry in the \pss{},
see \prettyref{fig:trapped_orbit_one_extent}, where this collection of
regular tori constitutes an effective surface.

It is important to emphasize that each stochastic
layer actually consists of a whole collection
of resonances. For example, the stochastic layer around
the ${22\!:\!0\!:\!7}$ resonance corresponds to the whole interval
with ${0.3173 \lesssim \nuone \lesssim 0.3186}$, see
\prettyref{fig:trapped_orbit_one_freq_space-all}(c).
This covers many resonance lines, see the points arranged on lines in
\prettyref{fig:trapped_orbit_one_freq_space-all}(a)
which correspond to higher order resonances.
Still, the ${22\!:\!0\!:\!7}$ resonance is the most dominant one
in this interval as the density of the frequency points $\nuone(t)$
is largest in its surrounding.

The sudden transitions between different frequency intervals are
manifestations of partial barriers.
For comparison this is illustrated in \prettyref{app:pb_2d}
for the \twoD{} billiard shown in \prettyref{fig:ps_2d_04}.
There, a sticky orbit approaches the boundary circle in the so-called
level hierarchy.
In such a two-dimensional case partial barriers
are well established as cantorus barriers (broken KAM curves)
or broken separatrices formed by stable and unstable manifolds \cite{Mei2015}.
However, these partial barriers do not generalize
to systems with more than two degrees of freedom.
Thus by using the frequency analysis it is possible
to detect partial barriers without constructing them explicitly,
in particular even if their dynamical origin is not known.

These results show that long-trapped orbits explore the chaotic part of
resonance channels and jump
(iii.a) across resonances, i.e.\ trapping takes place in the Chirikov
regime of overlapping resonances.
We find both in frequency space and in phase space clear signatures
of some kind of partial transport barriers.
At present their dynamical origin is not known.

\subsubsection*{(iii.b) Along resonance channels}

Besides the transport across resonance channels also
transport along resonance channels is present.
This is best visible for the considered long-trapped orbit
around the ${24\!:\!-2\!:\!7}$ resonance, as shown in the magnification
in \prettyref{fig:trapped_orbit_one_freq_space-all}(d).
Around the resonance line there is a characteristic
triangular-shaped region which is void of any regular \twoD{}
tori \cite{OnkLanKetBae2016}.
The extent is smallest near the Lyapunov family \Mone{}
and widens for increasing $\tilde{\nu}_y$.
As discussed in Sec.~\ref{sec:channels-AD} this
corresponds to going from the Nekhoroshev regime,
where the channel is surrounded by many regular tori
and transport is governed by very slow Arnold diffusion,
towards the Chirikov regime of overlapping resonances
for which the regular tori are sparse or not present.
Thus the distance to \Mone{} along the resonance channel 
takes the role of the perturbation strength 
in the setting of perturbed integrable dynamics.
While individual \twoD{} tori in a \fourD{} map
cannot confine chaotic motion,
a two-parameter family of them (with small gaps due to higher-order resonances)
effectively confines the chaotic motion around
the resonance within the triangular region in frequency space.

During the time interval $[7.5 \times 10^{8}, 1.1 \times 10^{9}]$
the orbit is located in the stochastic layer around
the ${24\!:\!-2\!:\!7}$ resonance,
i.e.\ $ \nuone{} \in \left[0.315, 0.3155 \right]$.
In the adapted coordinate $\tilde{\nu}_y$ one can
see that it initially decreases, i.e.\ the
sticky orbit moves along the channel towards
the Lyapunov family \Mone{},
see \prettyref{fig:trapped_orbit_one_freq_space-all}(d).
This approach is followed by
a longer time-interval with fluctuations
around some constant $\tilde{\nu}_y$
before the orbit moves along the channel
away from the Lyapunov family \Mone{}.
The involved time-scales show that the
motion along the resonance channel is typically much slower
than the motion within the stochastic layer, i.e.
see spreading in \nuone-direction or animation of
\prettyref{fig:trapped_orbit_one_extent}.

The numerical results indicate that the transition rates
for going across resonance channels
depend on the position along a resonance channel,
see \prettyref{fig:trapped_orbit_one_freq_space-all}(d).
In the considered time interval the adapted frequency
$\tilde{\nu}_y$ of the long-trapped orbit
first decreases, interpreted as approaching the Nekhoroshev regime in
which transitions across resonance channels become unlikely.
Subsequently the orbit moves along the channel
away from \Mone{} into the Chirikov regime allowing
for transitions across resonance channels.
Note that in Ref.~\cite{LanBaeKet2016}
it is suggested that transport along channels can be modeled by
a stochastic process with effective drift which gives one possible
mechanism of power-law trapping.

Since the Arnold web of connected resonance channels is
not explored on the considered time scales,
Arnold diffusion is not the origin of the long-trapped orbits.

\section{Summary and outlook}\label{sec:summary}

In this paper we visualize the mixed \ps{} of a \threeD{} billiard and
analyze restricted classical transport, manifested by
a slow decay of the Poincar\'e recurrence statistics, due to
long-trapped orbits.
To understand the dynamics of a \threeD{} billiard
its \sixD{} \ps{} is reduced by energy conservation and a
 \poinsec{} to a \fourD{} symplectic map.
This \fourD{} map with mixed \ps{} is visualized using a \pss{}
which reduces the dimension of orbits and invariant objects
such that they can be displayed in a \threeD{} plot.
This provides a good overview of the regular region and
its organization.
A complementary representation is the \twoD{} frequency space
in which both regular tori and sticky trajectories can be represented
and related to resonances.
Moreover, the frequency computation provides a chaos indicator to
distinguish between regular and chaotic dynamics.
The orbits in the \pss{} and in frequency space can be related to
trajectories in configuration space of the \threeD{} billiard
which provides an instructive
representation of objects in higher-dimensional systems.

The second focus of the paper is to study transport properties.
A slow power-law decay of the \poincare{} recurrence statistics
indicates the presence of sticky orbits.
This is of particular interest
as the mechanism of stickiness
for higher-dimensional systems is still not understood,
in contrast to trapping in systems with two degrees of freedom.
By analyzing long-trapped orbits in the \pss{} and in frequency space we
find that trapping takes place
(i) at the ''surface'' of the regular region (outside the Arnold web)
and is (ii) not due to a generalized island-around-island hierarchy.
We find that the dynamics of long-trapped orbits is (iii)
governed by numerous resonance channels which extend far into the chaotic sea.
The sticky orbits stay for long times within a stochastic layer
of a resonance channel, with fast transitions to other channels.
These are clear signatures, that between the stochastic layers
there are some restrictive partial barriers, whose
dynamical origin is not yet clear.
For the \threeD{} billiards the results in the \pss{},
see \prettyref{fig:trapped_orbit_one_extent} in particular,
suggest the existence of an effective (local) boundary surface
formed by regular \twoD{} tori
which is approached by the sticky chaotic orbits
via a sequence of coupled and uncoupled resonances.

An important task for the future is to identify and compute
the relevant partial barriers. Based on this it should
be possible to define the different states
and ultimately explain the origin of power-law trapping in
higher-dimensional systems.
Another interesting application of the visualization
of the phase space of a \threeD{} billiard
are \threeD{} optical microcavities
where understanding the mixed phase space may guide how
to tune their emission patterns.

\begin{acknowledgments}
  We are grateful for discussions with Swetamber Das, Felix Fritzsch,
  Franziska Onken, Martin Langer, Martin Richter, and Tom Schilling.
  Furthermore, we acknowledge support by the Deutsche Forschungsgemeinschaft
  under grant KE~537/6--1.

  All \threeDD{} visualizations were created using
  \textsc{Mayavi}~\cite{RamVar2011}.

\end{acknowledgments}

\appendix

\section{Signatures of partial barriers in 2D billiards}\label{app:pb_2d}

\begin{figure}[b]
    \includegraphics{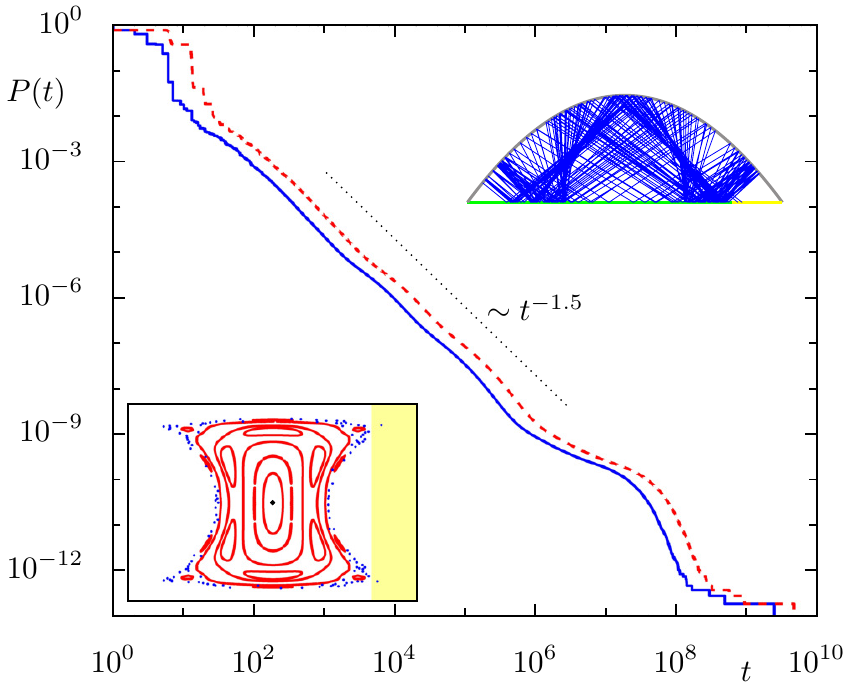}
    \caption{\label{fig:recurrence_2D}%
        \Prs{} $P(t)$ for the \twoD{} billiard with $a=1.04$
        for \(10^{13}\) trajectories started in region
        \(\Lambda\)
        for the number of mappings (blue line)
        and real flight time (red dashed line).
        The dotted line indicates a \pol{} decay \( \sim t^{-\gamma} \)
        with $\gamma = 1.5$.
        Upper inset: opened \twoD{} billiard with parabola as boundary
        and a short sticky trajectory with ${t_{\text{rec}} = 131}$.
        Lower inset: Poincar\'e section $(x, p_x)$
        with regular tori (red curves), opening \initregion{}
        (yellow rectangle), and
        corresponding trapped orbit (blue dots).}
\end{figure}

In \twoD{} billiards, and more generally in
autonomous Hamiltonian systems with two degrees-of-freedom, the
origin for power-law trapping are partial
transport barriers \cite{Mei1992,Mei2015}.
We now illustrate how signatures of these partial barriers can be detected in
the frequency-time plots for a \twoD{} billiard.
This allows for comparing with the corresponding time-frequency plots
of the \threeD{} billiard.
As an example we consider the \twoD{} billiard shown in
\prettyref{fig:ps_2d_04}
and determine the Poincar\'e recurrence statistics
as in  \prettyref{sec:prs}.
The result in \prettyref{fig:recurrence_2D}
shows an overall \pol{} with exponent $\gamma \approx 1.5$.
The slower decay around $t\approx 10^7$ is presumably
caused by some more restrictive partial barriers.

\begin{figure}[t]
    \includegraphics{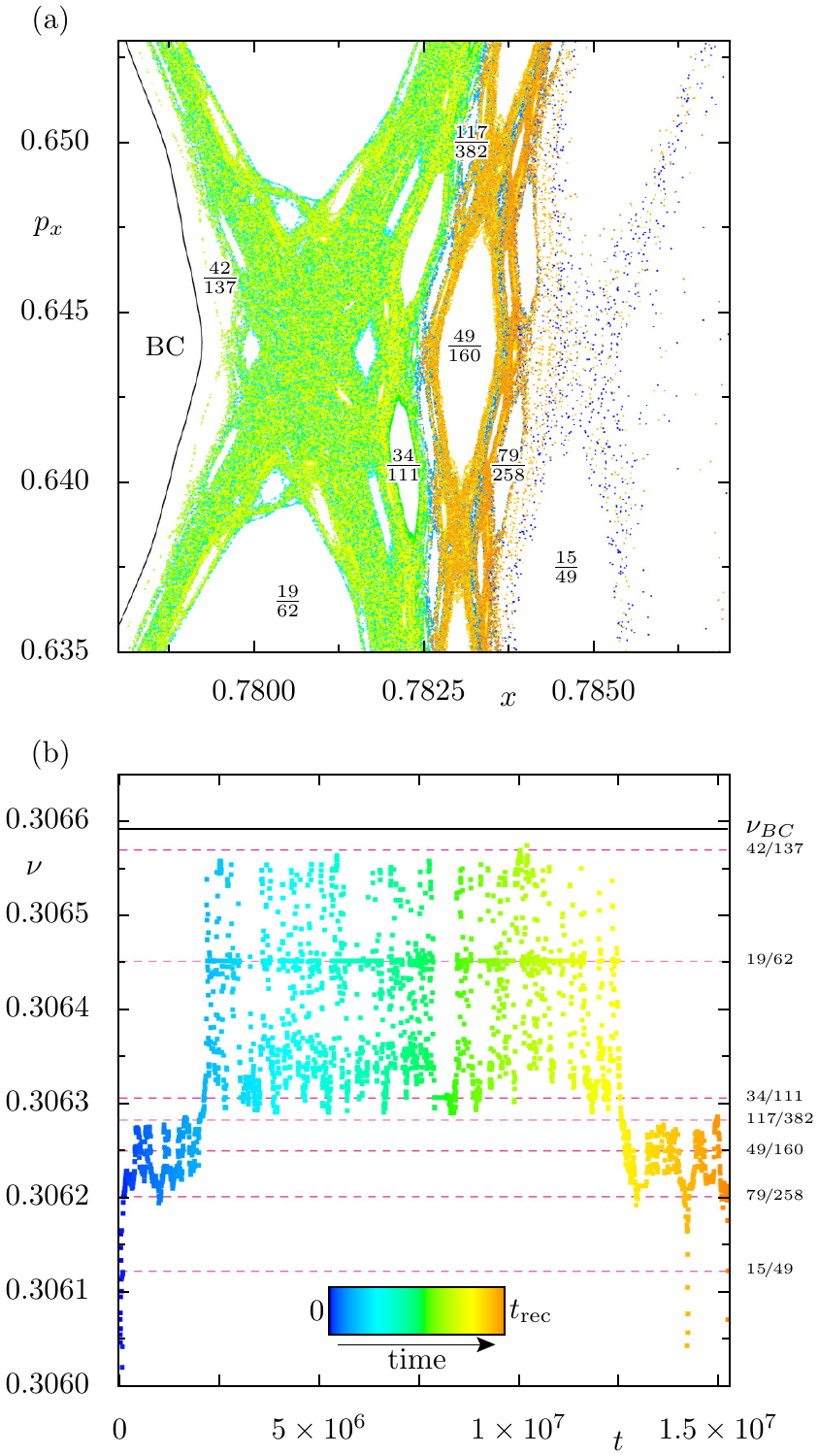}
    \caption{\label{fig:freq_time_sticky_2d}%
     (a)  Trapped orbit of the \twoD{} billiard with $a=1.04$ with
     recurrence time
      $t_{\text{rec}}\simeq1.52 \times 10^7$ in a magnification of \ps{} with
     time $[ 0, t_{\text{rec}} )$ encoded in color. Additionally shown is
     the boundary circle (full, black curve)
     and important surrounding resonances are labeled as fraction $n/m$.
     For the time-resolved animation see
     \MOVIELINK{}.
     (b) Frequency-time representation $\nu(t)$ of the sticky orbit.
     Frequencies of important resonances are shown as
     magenta dashed lines and the frequency $\nu_{\text{BC}}$ of the boundary
     circle as solid horizontal line.
        }
\end{figure}

\prettyref{fig:freq_time_sticky_2d} shows a long-trapped
orbit with ${t_{\text{rec}}\simeq 1.52 \times 10^{7}}$
and time encoded by color (blue to orange) in
a magnification of phase space and in frequency-time representation $\nu(t)$,
computed as in section \prettyref{sec:long_trapped}
from the complex signal $x-\ui p_x$ for
segments of length \( N_{\text{seg}} = 4096 \).
In phase space the long-trapped orbit approaches the boundary circle
of the central regular island, depicted in
an animation of the time-evolution
provided in the Supplemental Material at \MOVIELINK{}.
The boundary circle is the last invariant curve
and thus dynamically separates the regular region from
the surrounding chaotic motion.
The boundary circle has an irrational frequency
$\nu_{\text{BC}}$ which can be approximated by
the convergents of its continued fraction expansion \cite{GreMacSta1986}.
For the boundary circle with frequency $\nu_{\text{BC}} \approx 0.30659$,
the first approximants
are $\frac{3}{10}$, $\frac{4}{13}$, $\frac{19}{62}$, $\frac{23}{75}$,
$\frac{42}{137}$. Note that only every second approximant
is smaller than $\nu_{\text{BC}}$, giving
the sequence of principal resonances
$\frac{3}{10}$, $\frac{19}{62}$, and $\frac{42}{137}$.
Each of these fractions corresponds
to a resonance chain with elliptic
periodic orbits, surrounded by regular motion,
and hyperbolic periodic orbits with associated
chaotic layer.
These chaotic layers correspond to the states in a Markov model
description of power-law trapping
and separated from each other by partial barriers.
These partial barriers are cantori, broken KAM tori,
with irrational frequency $\nu_{\text{c}}$ which themselves can be approximated
by periodic orbits corresponding to the convergents of the continued fraction
expansion of $\nu_{\text{c}}$.
The transition rates between the states corresponding
to the stochastic layers become smaller
and smaller when approaching the boundary circle.
A long-trapped orbit is expected to approach the boundary circle via
this so-called level hierarchy of such states.
Note that the stochastic component of each of these states
usually contains several
other (non-principal) resonances.
Moreover, trapping also takes place in the neighborhood of the
resonance islands and their island-around-island hierarchy
\cite{Mei1986, MeiOtt1985, WeiHufKet2003},
which leads to time-intervals with constant frequency.

The different stochastic layers correspond
to the regions with different colors
in \prettyref{fig:freq_time_sticky_2d}(a).
Signatures of the partial transport barriers can also be clearly
seen in the frequency-time plot. Here the signal $\nu (t)$
randomly fluctuates within some interval around
a principal resonant frequency.
Passing through a partial barrier, a different
frequency interval around another dominant resonance is accessed.
For the example shown in \prettyref{fig:freq_time_sticky_2d}(b)
the frequencies of the sticky orbit are
initially confined in an interval around
$\nu = \frac{n}{m} = \frac{49}{160}$ and
then a sudden transition to the stochastic layer around
$\nu = \frac{19}{62}$ occurs.
This is one of the convergents of $\nu_{\text{BC}}$
and is closer to the boundary circle, compare
with \prettyref{fig:freq_time_sticky_2d}(a).
The stochastic layer around the next convergent $\nu = \frac{42}{137}$ is
only accessed very briefly.
Finally the level-hierarchy is left
via the stochastic layer around $\nu = \frac{49}{160}$ and by passing
through $\nu = \frac{15}{49}$ and $\nu = \frac{3}{10}$ (not shown).
Note that in this example not only stochastic regions
associated with the convergents of the boundary circle,
but also several other non-principal
resonances and partial barriers appear to be of relevance
for the long-time stickiness of the trapped orbit.

\section{Further example of a trapped orbit
           in the 3D billiard}
\label{app:trapped-2}

\begin{figure*}
        \includegraphics{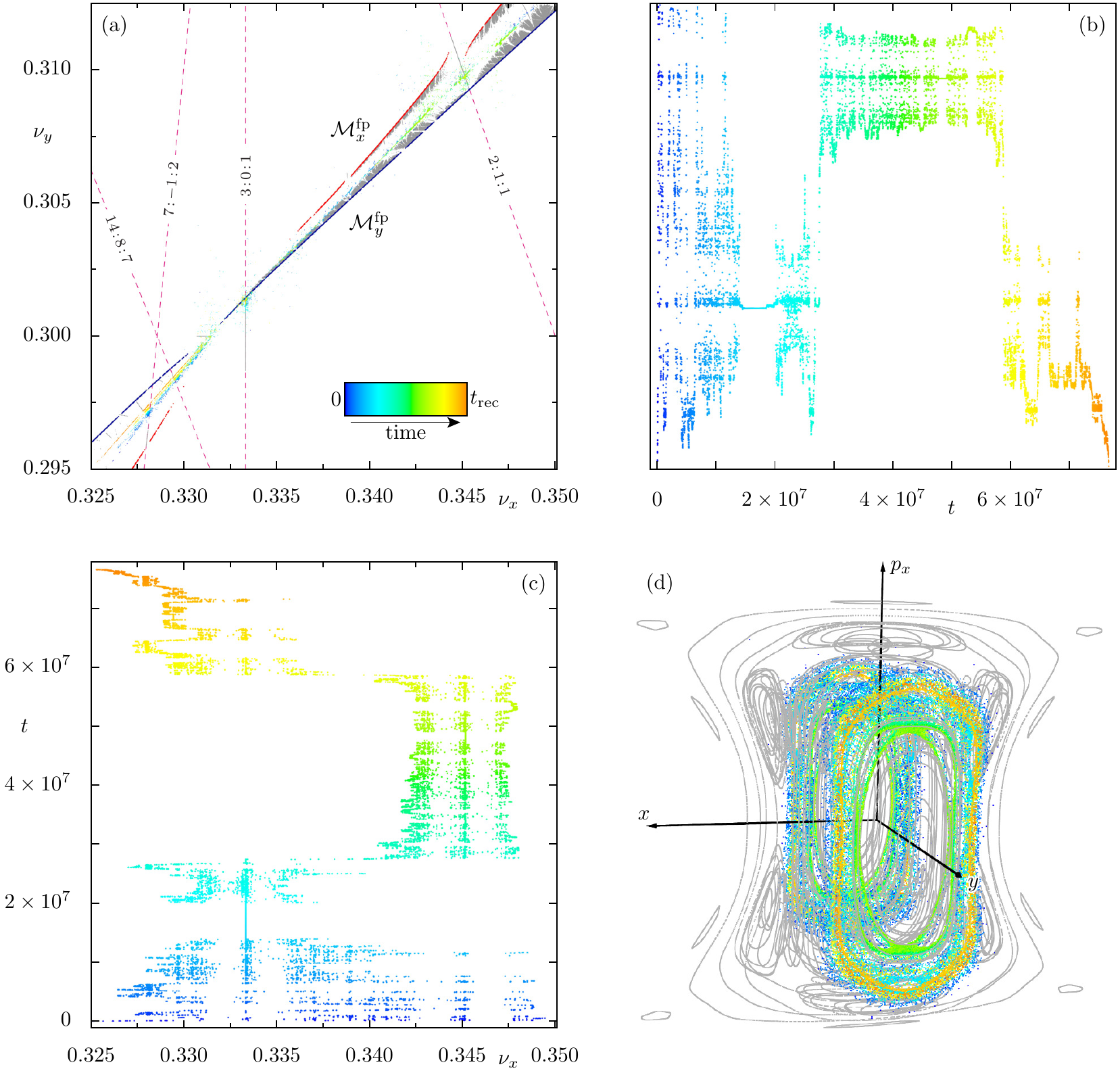}
    \caption{\label{fig:trapped_orbit_two_freq_space}
        Trapped orbit with recurrence time
        $t_{\text{rec}}\simeq 7.6 \times 10^{7}$ and
        time $t$ encoded in color.
        (a) Magnification of \prettyref{fig:freq_space}
        with regular tori (grey dots) and selected resonance lines,
        (b, c) frequency-time signals \( \nutwo(t) \) and
        \( \nuone(t) \),
        (d) \pss{} representation, for
        a rotating view see \MOVIELINK{}.
    }
\end{figure*}

A second example of a long-trapped orbit in the \threeD{} paraboloid billiard
is shown in \prettyref{fig:trapped_orbit_two_freq_space}.
Both in the \pss{} and in frequency space, see
\prettyref{fig:trapped_orbit_two_freq_space}\mbox{(a, d)}, the orbit is located
in a different region than the example discussed in
\prettyref{sec:long_trapped},
Figs.~\ref{fig:trapped_ps_one}--\ref{fig:trapped_orbit_one_freq_space-all}.
In particular the plot in frequency space reveals that
the sticky orbit is close to the
region where \Mone{} has a gap due to the important ${3\!:\!0\!:\!1}$ resonance
and is contained between the stochastic layers of the
${2\!:\!1\!:\!1}$ and the ${7\!:\!1\!:\!2}$ resonance.
In the frequency-time plots,
\prettyref{fig:trapped_orbit_two_freq_space}(b,~c),
the orbit first, up to $t \approx 1.3 \times 10^{7}$,
rapidly jumps between the stochastic layers of these two resonances,
and then mainly stays in the stochastic layer of either of them, i.e.\
for $t \in [3 \times 10^{7}, 5.6 \times 10^{7}]$
around the ${2\!:\!1\!:\!1}$ resonance clearly visible as green and yellow
colored stochastic layers in the \pss{} representation in
\prettyref{fig:trapped_orbit_two_freq_space}(d).
This highlights the possibility that different stochastic layers may
either behave as one large stochastic layer or
two separate ones.

For this orbit also trapping deeper in the hierarchy
is found for the time interval $t \in [1.5 \times 10^{7}, 2.2 \times 10^{7}]$
for which in
\prettyref{fig:trapped_orbit_two_freq_space}(b, c)
the frequency-time signals
collapse onto the frequencies of the ${3\!:\!0\!:\!1}$ resonance.
Note, that the deeper hierarchy around the ${3\!:\!0\!:\!1}$
resonance is not visible in the chosen perspective.

Thus overall this second example
shows both in phase space and in frequency space
conceptually the same signatures as the one discussed
in \prettyref{sec:long_trapped}.


\begin{thebibliography}{100}
    \newcommand{\enquote}[1]{``#1''}
    \providecommand{\url}[1]{\texttt{#1}}
    \providecommand{\urlprefix}{URL }
    \providecommand{\eprint}[2][]{\url{#2}}
    
    \bibitem{KozTre1991}
    V.~V. Kozlov and D.~V. Treshch\H{e}v, \emph{Billiards. A genetic Introduction
        to the Dynamics of Systems with Impacts}, volume~89 of \emph{Translations of
        Mathematical Monographs}, American Mathematical Society, Providence, Rhode
    Island (1991).
    
    \bibitem{Tab1995}
    S.~Tabachnikov, \emph{Billiards}, Panoramas et Synth{\`e}ses 1, Soci\'et\'e
    Mathematique de France, Paris (1995).
    
    \bibitem{CheMar2006}
    N.~Chernov and R.~Markarian, \emph{Chaotic Billiards}, volume 127 of
    \emph{Mathematical Surveys and Monographs}, American Mathematical Society,
    Providence, Rhode Island (2006).
    
    \bibitem{Sin1970}
    {\relax Ya}.~G. Sinai, \emph{Dynamical systems with elastic reflections},
    Russ.~Math.~Surv. \textbf{25}, 137 (1970).
    
    \bibitem{Bun1979}
    L.~A. Bunimovich, \emph{On the ergodic properties of nowhere dispersing
        billiards}, Commun.~Math.~Phys. \textbf{65}, 295 (1979).
    
    \bibitem{Rob1983}
    M.~Robnik, \emph{Classical dynamics of a family of billiards with analytic
        boundaries}, J.~Phys.~A \textbf{16}, 3971 (1983).
    
    \bibitem{Woj1986}
    M.~Wojtkowski, \emph{Principles for the design of billiards with nonvanishing
        {L}yapunov exponents}, Commun.~Math.~Phys. \textbf{105}, 391 (1986).
    
    \bibitem{Mar1988}
    R.~Markarian, \emph{Billiards with {Pesin} region of measure one},
    Commun.~Math.~Phys. \textbf{118}, 87 (1988).
    
    \bibitem{BaeDul1997}
    A.~B\"acker and H.~R. Dullin, \emph{Symbolic dynamics and periodic orbits for
        the cardioid billiard}, J.~Phys.~A \textbf{30}, 1991 (1997).
    
    \bibitem{MarMey1974}
    L.~Markus and K.~R. Meyer, \emph{Generic Hamiltonian Dynamical Systems are
        neither Integrable nor Ergodic}, number 144 in Mem.~Amer.~Math.~Soc.,
    American Mathematical Society, Providence, Rhode Island (1974).
    
    \bibitem{DulBae2001}
    H.~R. Dullin and A.~B\"acker, \emph{About ergodicity in the family of
        lima\c{c}on billiards}, Nonlinearity \textbf{14}, 1673 (2001).
    
    \bibitem{Bun2001}
    L.~A. Bunimovich, \emph{Mushrooms and other billiards with divided phase
        space}, Chaos \textbf{11}, 802 (2001).
    
    \bibitem{Sto2007b}
    H.-J. St\"ockmann, \emph{Quantum Chaos: {A}n {I}ntroduction}, Cambridge
    University Press, Cambridge (2007).
    
    \bibitem{Haa2010}
    F.~Haake, \emph{Quantum Signatures of Chaos}, Springer-Verlag, Berlin, 3rd
    {r}evised and {e}nlarged edition (2010).
    
    \bibitem{GmaCapNarNoeStoFaiSivCho1998}
    C.~Gmachl, F.~Capasso, E.~E. Narimanov, J.~U. N{\"o}ckel, A.~D. Stone,
    J.~Faist, D.~L. Sivco, and A.~Y. Cho, \emph{High-power directional emission
        from microlasers with chaotic resonators}, Science \textbf{280}, 1556 (1998).
    
    \bibitem{CaoWie2015}
    H.~Cao and J.~Wiersig, \emph{Dielectric microcavities: Model systems for wave
        chaos and non-{Hermitian} physics}, Rev.~Mod.~Phys. \textbf{87}, 61 (2015).
    
    \bibitem{Bun1988}
    L.~A. Bunimovich, \emph{Many-dimensional nowhere dispersing billiards with
        chaotic behavior}, Physica~D \textbf{33}, 58 (1988).
    
    \bibitem{Woj1990}
    M.~P. Wojtkowski, \emph{Linearly stable orbits in 3 dimensional billiards},
    Commun.~Math.~Phys. \textbf{129}, 319 (1990).
    
    \bibitem{BunCasGua1996}
    L.~A. Bunimovich, G.~Casati, and I.~Guarneri, \emph{Chaotic focusing billiards
        in higher dimensions}, Phys.~Rev.~Lett. \textbf{77}, 2941 (1996).
    
    \bibitem{BunReh1997}
    L.~A. Bunimovich and J.~Rehacek, \emph{Nowhere dispersing {3D} billiards with
        non-vanishing {Lyapunov} exponents}, Commun.~Math.~Phys. \textbf{189}, 729
    (1997).
    
    \bibitem{BunReh1998}
    L.~A. Bunimovich and J.~Rehacek, \emph{On the ergodicity of many-dimensional
        focusing billiards}, Ann. Inst. Henri Poincar\'e A \textbf{68}, 421 (1998).
    
    \bibitem{BunReh1998b}
    L.~A. Bunimovich and J.~Rehacek, \emph{How high-dimensional stadia look like},
    Commun.~Math.~Phys. \textbf{197}, 277 (1998).
    
    \bibitem{Bun2000}
    L.~A. Bunimovich, \emph{Hyperbolicity and astigmatism}, J.~Stat.~Phys.
    \textbf{101}, 373 (2000).
    
    \bibitem{BunMag2006}
    L.~A. Bunimovich and G.~Del~Magno, \emph{Semi-focusing billiards:
        Hyperbolicity}, Commun.~Math.~Phys. \textbf{262}, 17 (2006).
    
    \bibitem{Woj2007}
    M.~P. Wojtkowski, \emph{Design of hyperbolic billiards}, Commun.~Math.~Phys.
    \textbf{273}, 283 (2007).
    
    \bibitem{RapRomTur2007}
    A.~Rapoport, V.~Rom-Kedar, and D.~Turaev, \emph{Approximating multi-dimensional
        {Hamiltonian} flows by billiards}, Commun.~Math.~Phys. \textbf{272}, 567
    (2007).
    
    \bibitem{Sza2017}
    D.~Sz{\'a}sz, \emph{Multidimensional hyperbolic billiards}, in A.~M. Blokh,
    L.~A. Bunimovich, P.~H. Jung, L.~G. Oversteegen, and Y.~G. Sinai (editors)
    \enquote{Dynamical Systems, Ergodic Theory, and Probability: in Memory of
        Kolya Chernov}, volume 698 of \emph{Contemporary Mathematics}, American
    Mathematical Society, Providence, Rhode Island (2017).
    
    \bibitem{ZasStr1992}
    G.~M. Zaslavsky and H.~R. Strauss, \emph{Billiard in a barrel}, Chaos
    \textbf{2}, 469 (1992).
    
    \bibitem{PriSmi1995}
    H.~Primack and U.~Smilansky, \emph{Quantization of the three-dimensional
        {S}inai billiard}, Phys.~Rev.~Lett. \textbf{74}, 4831 (1995).
    
    \bibitem{PriSmi2000}
    H.~Primack and U.~Smilansky, \emph{The quantum three-dimensional {S}inai
        billiard -- a semiclassical analysis}, Phys.~Rep. \textbf{327}, 1 (2000).
    
    \bibitem{AltGraHofRanRehRicSchWir1996}
    H.~Alt, H.-D. Gr\"af, R.~Hofferbert, C.~Rangacharyulu, H.~Rehfeld, A.~Richter,
    P.~Schardt, and A.~Wirzba, \emph{Chaotic dynamics in a three-dimensional
        superconducting microwave billiard}, Phys.~Rev.~E \textbf{54}, 2303 (1996).
    
    \bibitem{Pro1997a}
    T.~Prosen, \emph{Quantization of generic chaotic 3d billiard with smooth
        boundary {I}. {E}nergy level statistics}, Phys.~Lett.~A \textbf{233}, 323
    (1997).
    
    \bibitem{Pro1997b}
    T.~Prosen, \emph{Quantization of generic chaotic {3D} billiard with smooth
        boundary {II}: {S}tructure of high-lying eigenstates}, Phys.~Lett.~A
    \textbf{233}, 332 (1997).
    
    \bibitem{Sie1998}
    M.~Sieber, \emph{Billiard systems in three dimensions: the boundary integral
        equation and the trace formula}, Nonlinearity \textbf{11}, 1607 (1998).
    
    \bibitem{Kni1998}
    O.~Knill, \emph{On nonconvex caustics of convex billiards}, Elem. Math.
    \textbf{53}, 89 (1998).
    
    \bibitem{WaaWieDul1999}
    H.~Waalkens, J.~Wiersig, and H.~R. Dullin, \emph{Triaxial ellipsoidal quantum
        billiards}, Ann.~Phys. \textbf{276}, 64 (1999).
    
    \bibitem{Pap2000}
    T.~Papenbrock, \emph{Numerical study of a three-dimensional generalized stadium
        billiard}, Phys.~Rev.~E \textbf{61}, 4626 (2000).
    
    \bibitem{Pap2000b}
    T.~Papenbrock, \emph{Lyapunov exponents and {Kolmogorov}-{Sinai} entropy for a
        high-dimensional convex billiard}, Phys.~Rev.~E \textbf{61}, 1337 (2000).
    
    \bibitem{PapPro2000}
    T.~Papenbrock and T.~Prosen, \emph{Quantization of a billiard model for
        interacting particles}, Phys.~Rev.~Lett. \textbf{84}, 262 (2000).
    
    \bibitem{DemDieGraHeiPapRicRic2002}
    C.~Dembowski, B.~Dietz, H.-D. Gr\"af, A.~Heine, T.~Papenbrock, A.~Richter, and
    C.~Richter, \emph{Experimental test of a trace formula for a chaotic
        three-dimensional microwave cavity}, Phys.~Rev.~Lett. \textbf{89}, 064101
    (2002).
    
    \bibitem{DieMoePapReiRic2008}
    B.~Dietz, B.~M\"o{\ss}ner, T.~Papenbrock, U.~Reif, and A.~Richter,
    \emph{Bouncing ball orbits and symmetry breaking effects in a
        three-dimensional chaotic billiard}, Phys.~Rev.~E \textbf{77}, 046221 (2008).
    
    \bibitem{CasRam2011}
    P.~Casas and R.~Ram{\'\i}rez-Ros, \emph{The frequency map for billiards inside
        ellipsoids}, {SIAM} J.~Appl.~Dyn.~Syst. \textbf{10}, 278 (2011).
    
    \bibitem{GilSan2011}
    T.~Gilbert and D.~P. Sanders, \emph{Stable and unstable regimes in
        higher-dimensional convex billiards with cylindrical shape}, New J. Phys.
    \textbf{13}, 023040 (2011).
    
    \bibitem{MekNoeCheStoCha1995}
    A.~Mekis, J.~U. N{\"o}ckel, G.~Chen, A.~D. Stone, and R.~K. Chang, \emph{Ray
        chaos and ${Q}$ spoiling in lasing droplets}, Phys.~Rev.~Lett. \textbf{75},
    2682 (1995).
    
    \bibitem{RakYanMcCDonPerMooGapRog2003}
    Y.~P. Rakovich, L.~Yang, E.~M. McCabe, J.~F. Donegan, T.~Perova, A.~Moore,
    N.~Gaponik, and A.~Rogach, \emph{Whispering gallery mode emission from a
        composite system of {CdTe} nanocrystals and a spherical microcavity},
    Semicond.~Sci.~Technol. \textbf{18}, 914 (2003).
    
    \bibitem{ChiDumFerFerJesNunPelSorRig2010}
    A.~Chiasera, Y.~Dumeige, P.~F{\'e}ron, M.~Ferrari, Y.~Jestin, G.~Nunzi~Conti,
    S.~Pelli, S.~Soria, and G.~Righini, \emph{Spherical whispering-gallery-mode
        microresonators}, Laser Photonics Rev. \textbf{4}, 457 (2010).
    
    \bibitem{KreSinHen2017}
    J.~Kreismann, S.~Sinzinger, and M.~Hentschel, \emph{Three-dimensional
        lima\c{c}on: Properties and applications}, Phys.~Rev.~A \textbf{95},
    011801(R) (2017).
    
    \bibitem{Arn1964}
    V.~I. Arnol'd, \emph{Instability of dynamical systems with several degrees of
        freedom}, Sov.~Math.~Dokl. \textbf{5}, 581 (1964).
    
    \bibitem{Chi1979}
    B.~V. {Chirikov}, \emph{{A universal instability of many-dimensional oscillator
            systems}}, Phys.~Rep. \textbf{52}, 263 (1979).
    
    \bibitem{Loc1999}
    P.~Lochak, \emph{Arnold diffusion; {A} compendium of remarks and questions}, in
    C.~Sim{\'o} (editor) \enquote{{H}amiltonian {S}ystems with {T}hree or {M}ore
        {D}egrees of {F}reedom}, volume 533 of \emph{NATO ASI Series: C -
        Mathematical and Physical Sciences}, 168, Kluwer Academic Publishers,
    Dordrecht (1999).
    
    \bibitem{Dum2014}
    H.~S. Dumas, \emph{The KAM Story: A Friendly Introduction to the Content,
        History, and Significance of Classical {Kolmogorov}--{Arnold}--{Moser}
        Theory}, World Scientific, Singapore (2014).
    
    \bibitem{Bae2007c}
    A.~B\"{a}cker, \emph{Quantum chaos in billiards}, Computing in Science \&
    Engineering \textbf{9}, 60 (2007).
    
    \bibitem{RicLanBaeKet2014}
    M.~Richter, S.~Lange, A.~B\"acker, and R.~Ketzmerick, \emph{Visualization and
        comparison of classical structures and quantum states of four-dimensional
        maps}, Phys.~Rev.~E \textbf{89}, 022902 (2014).
    
    \bibitem{Fro1972}
    C.~{Froeschl\'e}, \emph{Numerical study of a four-dimensional mapping},
    Astron.~\&~Astrophys. \textbf{16}, 172 (1972).
    
    \bibitem{LanRicOnkBaeKet2014}
    S.~Lange, M.~Richter, F.~Onken, A.~B\"acker, and R.~Ketzmerick, \emph{Global
        structure of regular tori in a generic {4D} symplectic map}, Chaos
    \textbf{24}, 024409 (2014).
    
    \bibitem{OnkLanKetBae2016}
    F.~Onken, S.~Lange, R.~Ketzmerick, and A.~B\"acker, \emph{Bifurcations of
        families of {1D}-tori in {4D} symplectic maps}, Chaos \textbf{26}, 063124
    (2016).
    
    \bibitem{BauBer1990}
    W.~Bauer and G.~F. Bertsch, \emph{Decay of ordered and chaotic systems},
    Phys.~Rev.~Lett. \textbf{65}, 2213 (1990).
    
    \bibitem{ZasTip1991}
    G.~M. Zaslavsky and M.~K. Tippett, \emph{Connection between recurrence-time
        statistics and anomalous transport}, Phys.~Rev.~Lett. \textbf{67}, 3251
    (1991).
    
    \bibitem{HirSauVai1999}
    M.~Hirata, B.~Saussol, and S.~Vaienti, \emph{Statistics of return times: A
        general framework and new applications}, Commun.~Math.~Phys. \textbf{206}, 33
    (1999).
    
    \bibitem{AltSilCal2004}
    E.~G. Altmann, E.~C. {da Silva}, and I.~L. Caldas, \emph{Recurrence time
        statistics for finite size intervals}, Chaos \textbf{14}, 975 (2004).
    
    \bibitem{ChaLeb1980}
    S.~R. Channon and J.~L. Lebowitz, \emph{Numerical experiments in stochasticity
        and homoclinic oscillation}, Ann.~N.Y.~Acad.~Sci. \textbf{357}, 108 (1980).
    
    \bibitem{ChiShe1983}
    B.~V. Chirikov and D.~L. Shepelyansky, \emph{Statistics of {P}oincar\'e
        recurrences and the structure of the stochastic layer of a nonlinear
        resonance},  (1983), in \emph{Proceedings of the 9th International Conference
        on Nonlinear Oscillations, Kiev, 1981; Qualitative methods of the theory of
        nonlinear oscillators}, vol.~2, edited by Yu. A. Mitropolsky, 421--424,
    Naukova Dumka, Kiev, (1984). English translation: Princeton University Report
    No. PPPL-TRANS-133 (1983).
    
    \bibitem{Kar1983}
    C.~F.~F. Karney, \emph{Long-time correlations in the stochastic regime},
    Physica~D \textbf{8}, 360 (1983).
    
    \bibitem{ChiShe1984}
    B.~V. Chirikov and D.~L. Shepelyansky, \emph{Correlation properties of
        dynamical chaos in {Hamiltonian} systems}, Physica~D \textbf{13}, 395 (1984).
    
    \bibitem{KayMeiPer1984a}
    R.~S. MacKay, J.~D. Meiss, and I.~C. Percival, \emph{Stochasticity and
        transport in {Hamiltonian} systems}, Phys.~Rev.~Lett. \textbf{52}, 697
    (1984).
    
    \bibitem{HanCarMei1985}
    J.~D. Hanson, J.~R. Cary, and J.~D. Meiss, \emph{Algebraic decay in
        self-similar {Markov} chains}, J.~Stat.~Phys. \textbf{39}, 327 (1985).
    
    \bibitem{Mei1986}
    J.~D. Meiss, \emph{Class renormalization: Islands around islands}, Phys.~Rev.~A
    \textbf{34}, 2375 (1986).
    
    \bibitem{MeiOtt1985}
    J.~D. Meiss and E.~Ott, \emph{Markov-tree model of intrinsic transport in
        {H}amiltonian systems}, Phys.~Rev.~Lett. \textbf{55}, 2741 (1985).
    
    \bibitem{MeiOtt1986}
    J.~D. Meiss and E.~Ott, \emph{Markov tree model of transport in area-preserving
        maps}, Physica~D \textbf{20}, 387 (1986).
    
    \bibitem{ZasEdeNiy1997}
    G.~M. Zaslavsky, M.~Edelmann, and B.~A. Niyazov, \emph{Self-similarity,
        renormalization, and phase space nonuniformity of {Hamiltonian} chaotic
        dynamics}, Chaos \textbf{7}, 159 (1997).
    
    \bibitem{BenKasWhiZas1997}
    S.~Benkadda, S.~Kassibrakis, R.~B. White, and G.~M. Zaslavsky,
    \emph{Self-similarity and transport in the standard map}, Phys.~Rev.~E
    \textbf{55}, 4909 (1997).
    
    \bibitem{ChiShe1999}
    B.~V. Chirikov and D.~L. Shepelyansky, \emph{Asymptotic statistics of
        {Poincar\'e} recurrences in {Hamiltonian} systems with divided phase space},
    Phys.~Rev.~Lett. \textbf{82}, 528 (1999).
    
    \bibitem{ZasEde2000}
    G.~M. Zaslavsky and M.~Edelmann, \emph{Hierarchical structures in the phase
        space and fractional kinetics: {I}. {C}lassical systems}, Chaos \textbf{10},
    135 (2000).
    
    \bibitem{WeiHufKet2003}
    M.~Weiss, L.~Hufnagel, and R.~Ketzmerick, \emph{Can simple renormalization
        theories describe the trapping of chaotic trajectories in mixed systems?},
    Phys. Rev. E \textbf{67}, 046209 (2003).
    
    \bibitem{CriKet2008}
    G.~Cristadoro and R.~Ketzmerick, \emph{Universality of algebraic decays in
        {H}amiltonian systems}, Phys.~Rev.~Lett. \textbf{100}, 184101 (2008).
    
    \bibitem{CedAga2013}
    R.~Ceder and O.~Agam, \emph{Fluctuations in the relaxation dynamics of mixed
        chaotic systems}, Phys.~Rev.~E \textbf{87}, 012918 (2013).
    
    \bibitem{AluFisMei2014}
    O.~Alus, S.~Fishman, and J.~D. Meiss, \emph{Statistics of the
        island-around-island hierarchy in {H}amiltonian phase space}, Phys.~Rev.~E
    \textbf{90}, 062923 (2014).
    
    \bibitem{AluFisMei2017}
    O.~Alus, S.~Fishman, and J.~D. Meiss, \emph{Universal exponent for transport in
        mixed {H}amiltonian dynamics}, Phys.~Rev.~E \textbf{96}, 032204 (2017).
    
    \bibitem{DetMarStr2016}
    C.~P. Dettmann, J.~Marklof, and A.~Str{\"o}mbergsson, \emph{Universal hitting
        time statistics for integrable flows}, J.~Stat.~Phys. 1 (2016), vIP: use
    DetMarStr2017 as this is only the online published version.
    
    \bibitem{AltTel2008}
    E.~G. Altmann and T.~T{\'e}l, \emph{Poincar\'e recurrences from the perspective
        of transient chaos}, Phys.~Rev.~Lett. \textbf{100}, 174101 (2008).
    
    \bibitem{AltTel2009}
    E.~G. Altmann and T.~T{\'e}l, \emph{Poincar\'e recurrences and transient chaos
        in systems with leaks}, Phys.~Rev.~E \textbf{79}, 016204 (2009).
    
    \bibitem{AltPorTel2013}
    E.~G. Altmann, J.~S.~E. Portela, and T.~T\'el, \emph{Leaking chaotic systems},
    Rev.~Mod.~Phys. \textbf{85}, 869 (2013).
    
    \bibitem{DetRah2014}
    C.~P. Dettmann and M.~R. Rahman, \emph{Survival probability for open spherical
        billiards}, Chaos \textbf{24}, 043130 (2014).
    
    \bibitem{Aub1978}
    S.~Aubry, \emph{The new concept of transitions by breaking of analyticity in a
        crystallographic model}, in A.~R. Bishop and T.~Schneider (editors)
    \enquote{Solitons and {{Condensed Matter Physics}}}, volume~8 of
    \emph{Springer Series in Solid-State Sciences}, 264, {Springer Berlin
        Heidelberg} (1978).
    
    \bibitem{Per1980}
    I.~C. Percival, \emph{Variational principles for invariant tori and cantori},
    in \enquote{{{AIP Conference Proceedings}}}, volume~57, 302, {AIP Publishing}
    (1980).
    
    \bibitem{BenKad1984}
    D.~Bensimon and L.~P. Kadanoff, \emph{Extended chaos and disappearance of {KAM}
        trajectories}, Physica~D \textbf{13}, 82 (1984).
    
    \bibitem{KayMeiPer1984b}
    R.~S. MacKay, J.~D. Meiss, and I.~C. Percival, \emph{{Transport in
            {H}amiltonian systems}}, Physica~D \textbf{13}, 55 (1984).
    
    \bibitem{Wig1990}
    S.~Wiggins, \emph{On the geometry of transport in phase space {I}. {T}ransport
        in $k$-degree-of-freedom {H}amiltonian systems, $2 \leq k < \infty$},
    Physica~D \textbf{44}, 471 (1990).
    
    \bibitem{RomWig1990}
    V.~Rom-Kedar and S.~Wiggins, \emph{Transport in two-dimensional maps},
    Arch.~Rational Mech.~Anal. \textbf{109}, 239 (1990).
    
    \bibitem{RomWig1991}
    V.~Rom-Kedar and S.~Wiggins, \emph{Transport in two-dimensional maps: Concepts,
        examples, and a comparison of the theory of {Rom-Kedar} and {Wiggins} with
        the {Markov} model of {MacKay}, {Meiss}, {Ott}, and {Percival}}, Physica~D
    \textbf{51}, 248 (1991).
    
    \bibitem{Mei1992}
    J.~D. Meiss, \emph{Symplectic maps, variational principles, and transport},
    Rev.~Mod.~Phys. \textbf{64}, 795 (1992).
    
    \bibitem{Mei2015}
    J.~D. Meiss, \emph{Thirty years of turnstiles and transport}, Chaos
    \textbf{25}, 097602 (2015).
    
    \bibitem{GreMacSta1986}
    J.~M. Greene, R.~S. MacKay, and J.~Stark, \emph{Boundary circles for
        area-preserving maps}, Physica~D \textbf{21}, 267 (1986).
    
    \bibitem{KanKon1989}
    K.~Kaneko and T.~Konishi, \emph{Diffusion in {Hamiltonian} dynamical systems
        with many degrees of freedom}, Phys. Rev. A \textbf{40}, 6130 (1989).
    
    \bibitem{KonKan1990}
    T.~Konishi and K.~Kaneko, \emph{Diffusion in {H}amiltonian chaos and its size
        dependence}, J.~Phys.~A \textbf{23}, L715 (1990).
    
    \bibitem{DinBouOtt1990}
    M.~Ding, T.~Bountis, and E.~Ott, \emph{Algebraic escape in higher dimensional
        {Hamiltonian} systems}, Phys.~Lett.~A \textbf{151}, 395 (1990).
    
    \bibitem{ChiVec1993}
    B.~V. Chirikov and V.~V. Vecheslavov, \emph{Theory of fast {A}rnold diffusion
        in many-frequency systems}, J.~Stat.~Phys. \textbf{71}, 243 (1993).
    
    \bibitem{ChiVec1997}
    B.~V. Chirikov and V.~V. Vecheslavov, \emph{Arnold diffusion in large systems},
    J.~Exp.~Theor.~Phys. \textbf{85}, 616 (1997).
    
    \bibitem{AltKan2007}
    E.~G. Altmann and H.~Kantz, \emph{Hypothesis of strong chaos and anomalous
        diffusion in coupled symplectic maps}, Europhys.~Lett. \textbf{78}, 10008
    (2007).
    
    \bibitem{ShoLiKomTod2007b}
    A.~Shojiguchi, C.-B. Li, T.~Komatsuzaki, and M.~Toda, \emph{Fractional behavior
        in multidimensional {Hamiltonian} systems describing reactions}, Phys.~Rev.~E
    \textbf{76}, 056205 (2007), erratum ibid. {\bf 77}, 019902(E) (2008).
    
    \bibitem{She2010}
    D.~L. Shepelyansky, \emph{Poincar\'e recurrences in {Hamiltonian} systems with
        a few degrees of freedom}, Phys.~Rev.~E \textbf{82}, 055202(R) (2010).
    
    \bibitem{LanBaeKet2016}
    S.~Lange, A.~B{\"a}cker, and R.~Ketzmerick, \emph{What is the mechanism of
        power-law distributed {Poincar\'e} recurrences in higher-dimensional
        systems?}, EPL \textbf{116}, 30002 (2016).
    
    \bibitem{Bir1966}
    G.~D. Birkhoff, \emph{Dynamical Systems}, volume~9 of \emph{Colloquium
        publications}, American Mathematical Society, Providence, Rhode Island,
    revised edition (1966).
    
    \bibitem{Hig2002}
    N.~J. Higham, \emph{Accuracy and Stability of Numerical Algorithms}, Society
    for Industrial and Applied Mathematics, Philadelphia, PA, USA, second edition
    (2002).
    
    \bibitem{Poe2001}
    J.~P\"oschel, \emph{A lecture on the classical {KAM} theorem}, in A.~Katok,
    R.~de~la Llave, Y.~Pesin, and H.~Weiss (editors) \enquote{Smooth Ergodic
        Theory and its Applications}, volume~69 of \emph{Proceedings of Symposia in
        Pure Mathematics}, 707, Amer. Math. Soc., Providence, RI (2001).
    
    \bibitem{Lla2001}
    R.~de~la Llave, \emph{A tutorial on {KAM} theory}, in A.~Katok, R.~de~la Llave,
    Y.~Pesin, and H.~Weiss (editors) \enquote{Smooth Ergodic Theory and its
        Applications}, volume~69 of \emph{Proceedings of Symposia in Pure
        Mathematics}, 75–292, Amer. Math. Soc., Providence, RI (2001).
    
    \bibitem{Poi1912}
    H.~Poincar\'e, \emph{Sur un th\'eor\`eme de g\'eom\'etrie},
    Rend.~Cir.~Mat.~Palermo \textbf{33}, 375 (1912).
    
    \bibitem{Bir1913}
    G.~D. Birkhoff, \emph{Proof of {P}oincar\'e's geometric theorem},
    Trans.~Amer.~Math.~Soc. \textbf{14}, 14 (1913).
    
    \bibitem{AltMotKan2005}
    E.~G. Altmann, A.~E. Motter, and H.~Kantz, \emph{Stickiness in mushroom
        billiards}, Chaos \textbf{15}, 033105 (2005).
    
    \bibitem{TanShu2006}
    H.~Tanaka and A.~Shudo, \emph{Recurrence time distribution in mushroom
        billiards with parabolic hat}, Phys.~Rev.~E \textbf{74}, 036211 (2006).
    
    \bibitem{DetGeo2011}
    C.~P. Dettmann and O.~Georgiou, \emph{Open mushrooms: stickiness revisited},
    J.~Phys.~A \textbf{44}, 195102 (2011).
    
    \bibitem{BunVel2012}
    L.~A. Bunimovich and L.~V. Vela-Arevalo, \emph{Many faces of stickiness in
        {H}amiltonian systems}, Chaos \textbf{22}, 026103 (2012).
    
    \bibitem{DelGut1996}
    A.~Delshams and P.~Guti{\'e}rrez, \emph{Estimates on invariant tori near an
        elliptic equilibrium point of a {H}amiltonian system}, J.~Diff.~Eqs.
    \textbf{131}, 277 (1996).
    
    \bibitem{AnaBouBae2017}
    S.~Anastassiou, T.~Bountis, and A.~B{\"a}cker, \emph{Homoclinic points of {2D}
        and {4D} maps via the parametrization method}, Nonlinearity \textbf{30}, 3799
    (2017).
    
    \bibitem{JorVil1997}
    {\`{A}}.~Jorba and J.~Villanueva, \emph{On the normal behaviour of partially
        elliptic lower-dimensional tori of {Hamiltonian} systems}, Nonlinearity
    \textbf{10}, 783 (1997).
    
    \bibitem{JorVil1997b}
    {\`{A}}.~Jorba and J.~Villanueva, \emph{On the persistence of lower dimensional
        invariant tori under quasi-periodic perturbations}, J.~Nonlinear~Sci.
    \textbf{7}, 427 (1997).
    
    \bibitem{JorOll2004}
    {\`A}.~Jorba and M.~Oll{\'e}, \emph{Invariant curves near {Hamiltonian}-{Hopf}
        bifurcations of four-dimensional symplectic maps}, Nonlinearity \textbf{17},
    691 (2004).
    
    \bibitem{MarDavEzr1987}
    C.~C. Martens, M.~J. Davis, and G.~S. Ezra, \emph{{Local frequency analysis of
            chaotic motion in multidimensional systems: energy transport and bottlenecks
            in planar OCS}}, Chem.~Phys.~Lett. \textbf{142}, 519 (1987).
    
    \bibitem{Las1993}
    J.~Laskar, \emph{Frequency analysis for multi-dimensional systems. {G}lobal
        dynamics and diffusion}, Physica~D \textbf{67}, 257 (1993).
    
    \bibitem{BarBazGioScaTod1996}
    R.~Bartolini, A.~Bazzani, M.~Giovannozzi, W.~Scandale, and E.~Todesco,
    \emph{{{T}une evaluation in simulations and experiments}}, Part.~Accel.
    \textbf{52}, 147 (1996).
    
    \bibitem{DulMei2003}
    H.~R. Dullin and J.~D. Meiss, \emph{Twist singularities for symplectic maps},
    Chaos \textbf{13}, 1 (2003).
    
    \bibitem{GekMaiBarUze2007}
    S.~Gekle, J.~Main, T.~Bartsch, and T.~Uzer, \emph{Hydrogen atom in crossed
        electric and magnetic fields: Phase space topology and torus quantization via
        periodic orbits}, Phys.~Rev.~A \textbf{75}, 023406 (2007).
    
    \bibitem{SkoGotLas2016}
    {\relax Ch}.~Skokos, G.~A. Gottwald, and J.~Laskar (editors) \emph{Chaos
        Detection and Predictability}, volume 915 of \emph{Lecture Notes in Physics},
    {Springer Berlin Heidelberg} (2016).
    
    \bibitem{SieSte1990}
    M.~Sieber and F.~Steiner, \emph{Classical and quantum mechanics of a strongly
        chaotic billiard system}, Physica~D \textbf{44}, 248 (1990).
    
    \bibitem{CasJor2000}
    E.~Castell\'a and {\`A}.~Jorba, \emph{On the vertical families of
        two-dimensional tori near the triangular points of the bicircular problem},
    Celest.~Mech.~Dyn.~Astron. \textbf{76}, 35 (2000).
    
    \bibitem{Tod1994}
    E.~{Todesco}, \emph{{Analysis of resonant structures of four-dimensional
            symplectic mappings, using normal forms}}, Phys.~Rev.~E \textbf{50}, R4298
    (1994).
    
    \bibitem{Tod1996}
    E.~Todesco, \emph{Local analysis of formal stability and existence of fixed
        points in 4d symplectic mappings}, Physica~D \textbf{95}, 1 (1996).
    
    \bibitem{PatZac1994}
    P.~A. Patsis and L.~Zachilas, \emph{Using color and rotation for visualizing
        four-dimensional {P}oincar\'e cross-sections: With applications to the
        orbital behavior of a three-dimensional {H}amiltonian system},
    Int.~J.~Bifurcation~Chaos \textbf{4}, 1399 (1994).
    
    \bibitem{KatPat2011}
    M.~Katsanikas and P.~A. Patsis, \emph{The structure of invariant tori in a {3D}
        galactic potential}, Int.~J.~Bifurcation~Chaos \textbf{21}, 467 (2011).
    
    \bibitem{JorVil2001}
    {\`{A}}.~Jorba and J.~Villanueva, \emph{The fine geometry of the {Cantor}
        families of invariant tori in {Hamiltonian} systems}, in C.~Casacuberta,
    R.~Mir\'{o}-Roig, J.~Verdera, and S.~Xamb\'{o}-Descamps (editors)
    \enquote{European Congress of Mathematics}, volume 202 of \emph{Progress in
        Mathematics}, 557, Birkh\"auser Basel (2001).
    
    \bibitem{Nek1977}
    N.~N. Nekhoroshev, \emph{An exponential estimate of the time of stability of
        nearly-integrable {H}amiltonian systems}, Russ.~Math.~Surv. \textbf{32}, 1
    (1977).
    
    \bibitem{Guz2004}
    M.~Guzzo, \emph{A direct proof of the {N}ekhoroshev theorem for nearly
        integrable symplectic maps}, Ann.~Henri Poincar\'{e} \textbf{5}, 1013 (2004).
    
    \bibitem{GuzLegFro2002}
    M.~Guzzo, E.~Lega, and C.~{Froeschl{\'e}}, \emph{On the numerical detection of
        the effective stability of chaotic motions in quasi-integrable systems},
    Physica~D \textbf{163}, 1 (2002).
    
    \bibitem{FroLegGuz2006}
    C.~{Froeschl{\'e}}, E.~{Lega}, and M.~{Guzzo}, \emph{{Analysis of the Chaotic
            Behaviour of Orbits Diffusing along the {Arnold} Web}},
    Celest.~Mech.~Dyn.~Astron. \textbf{95}, 141 (2006).
    
    \bibitem{Poi1890}
    H.~{Poincar\'e}, \emph{Sur le probl\`eme des trois corps et les \'equations de
        la dynamique}, Acta~Math. \textbf{13}, 1 (1890).
    
    \bibitem{Kac1947}
    M.~Kac, \emph{On the notion of recurrence in discrete stochastic processes},
    Bulletin of the American Mathematical Society \textbf{53}, 1002 (1947).
    
    \bibitem{Kac1959}
    M.~Kac, \emph{Probability and Related Topics in Physical Sciences}, Lectures in
    Applied Mathematics Series, Vol 1a, American Mathematical Society (1959).
    
    \bibitem{Mei1997}
    J.~D. Meiss, \emph{Average exit time for volume-preserving maps}, Chaos
    \textbf{7}, 139 (1997).
    
    \bibitem{WeiHufKet2002}
    M.~Weiss, L.~Hufnagel, and R.~Ketzmerick, \emph{Universal power-law decay in
        {Hamiltonian} systems?}, Phys.~Rev.~Lett. \textbf{89}, 239401 (2002).
    
    \bibitem{AluFis2015}
    O.~Alus and S.~Fishman, \emph{Diffusion for ensembles of standard maps},
    Phys.~Rev.~E \textbf{92}, 042904 (2015).
    
    \bibitem{Fir2014}
    M.~Firmbach, \emph{3D Billards: Dynamik im gemischten Phasenraum und
        Potenzgesetz des H\"angenbleibens}, Bachelor thesis, Technische Universit\"at
    Dresden, Fachrichtung Physik (2014).
    
    \bibitem{ConHarVog2000}
    G.~Contopoulos, M.~Harsoula, and N.~Voglis, \emph{Crossing of various cantori},
    Celest.~Mech.~Dyn.~Astron. \textbf{78}, 197 (2000).
    
    \bibitem{Lan2016}
    S.~Lange, \emph{Chaotic transport and trapping close to regular structures in
        4D symplectic maps}, {Ph.D.} thesis, Technische Universit\"at Dresden,
    Fachrichtung Physik (2016).
    
    \bibitem{RamVar2011}
    P.~Ramachandran and G.~Varoquaux, \emph{{Mayavi}: {3D} visualization of
        scientific data}, Comput.~Sci.~Eng. \textbf{13}, 40 (2011).
    
\end{thebibliography}
\end{document}